\documentclass[%
 reprint,
nofootinbib,
 amsmath,amssymb,
 aps,
]{revtex4-1}
\usepackage{multirow}
\usepackage{adjustbox,lipsum}
\usepackage{amsmath}
\usepackage{cancel}
\usepackage{graphicx}
\usepackage{dcolumn}
\usepackage{bm}
\usepackage{tabularx}
\usepackage{xcolor}
\usepackage[colorlinks=true]{hyperref}
\usepackage{color}
\usepackage[normalem]{ulem}
\usepackage{url}
\usepackage[title]{appendix}
\newcommand\be{\begin{equation}}
\newcommand\ba{\begin{eqnarray}}
\newcommand\ee{\end{equation}}
\newcommand\ea{\end{eqnarray}}
\newcommand\bw{\begin{widetext}}
\newcommand\ew{\end{widetext}}

\newcommand{\ET}{{\mbox{\tiny (ET)}}}
\newcommand{\DECIGO}{{\mbox{\tiny (DEC)}}}
\newcommand{\ISCO}{{\mbox{\tiny ISCO}}}
\newcommand{\DE}{{\mbox{\tiny DE}}}

\begin{document}


\title{Probing modified gravitational-wave propagation through tidal measurements \\ of binary neutron star mergers}
\author{Nan Jiang}
\author{Kent Yagi}
\affiliation{Department of Physics, University of Virginia, Charlottesville, Virginia 22904, USA}

\date{\today}

\begin{abstract}

Gravitational-wave sources can serve as standard sirens to probe cosmology by measuring their luminosity distance and redshift.
Such standard sirens are also useful to probe theories beyond general relativity with a modified gravitational-wave propagation.
Many of previous studies on the latter assume multi-messenger observations so that the luminosity distance can be measured with gravitational waves while the redshift is obtained by identifying sources' host galaxies from electromagnetic counterparts. 
Given that gravitational-wave events of binary neutron star coalescences with associated electromagnetic counterpart detections are expected to be rather rare, it is important to examine the possibility of using standard sirens with gravitational-wave observations alone to probe gravity.
In this paper, we achieve this by extracting the redshift from the tidal measurement of binary neutron stars that was originally proposed within the context of gravitational-wave cosmology (another approach is to correlate ``dark sirens'' with galaxy catalogs that we do not consider here). 
We consider not only observations with ground-based detectors (e.g. Einstein Telescope) but also multi-band observations between ground-based and space-based (e.g. DECIGO) interferometers. 
We find that such multi-band observations with the tidal information can constrain a parametric non-Einsteinian deviation in the luminosity distance (due to the modified friction in the gravitational wave evolution) more stringently than the case with electromagnetic counterparts by a factor of a few.
We also map the above-projected constraints on the parametric deviation to those on specific theories and phenomenological models beyond general relativity to put the former into context.

\end{abstract}

\maketitle

\section{Introduction}
\label{sec:intro}

A historic detection of gravitational waves (GWs) was made September 14, 2015, by the Laser Interferometer Gravitational-wave Observatory (LIGO) in Hanford and Livingston. The GW event is known as GW150914~\cite{PhysRevLett.116.061102} and consists of a merger of a binary black hole (BBH). So far, nearly 50 BBH merger GW events have been found~\cite{Abbott:2020niy}. Another milestone observation was made in 2017 when LIGO and Virgo detected GW signals from a coalescing binary neutron star (BNS), known as GW170817~\cite{PhysRevLett.119.161101}. This event marked the dawn of multi-messenger astronomy as not only GW signals but also their associated electromagnetic (EM) counterparts were detected~\cite{Monitor:2017mdv}. A second BNS event, GW190425~\cite{Abbott_2020}, was found in the third observing run by the LIGO/Virgo Collaboration (LVC), though no electromagnetic counterpart is confirmed yet. 

GW170817 serves as a standard siren to probe cosmology, in particular measuring the Hubble constant~\cite{2017standardsiren,Fishbach_2019,Chen_2018}. This constant is inferred from the independent measurement of the luminosity distance and the redshift of the source. The former is measured from the GW amplitude while the latter is obtained by identifying the host galaxy through EM counterpart observations.

Another important application of GW170817 is to test general relativity (GR). Going beyond GR is motivated by the unification of GR and the Standard Model~\cite{Joyce_2015,Clifton_2012,nastase2012introduction}, and it can also explain some of unsolved problems in cosmology, such as dark matter and dark energy problems~\cite{Jain_2010,Salvatelli_2016,1983ApJ...270..365M,Famaey_2012,Belgacem_2019}. GR has passed all the tests put to it, including solar system experiments~\cite{Will_2014} in the weak-field regime, binary pulsar observations~\cite{Stairs_2003,wex2014testing} in the strong/non-dynamical regime and GW observations~\cite{TheLIGOScientific:2016src,PhysRevD.94.084002,Berti_2018,theligoscientificcollaboration2020tests, carson2021testing} in the strong/dynamical regime. GW170817 has been used to probe the modified dispersion relation of GWs. For example, the comparison of the arrival time difference between GW and EM wave signals placed a bound on the fractional difference in the propagation speed of GWs with respect to the speed of light to be one part in $10^{-15}$~\cite{Monitor:2017mdv}.

Standard sirens like GW170817 can also probe other aspects of the modified GW propagation, in particular the modified friction term in the GW evolution. This in turn modifies the GW amplitude from its GR counterpart, and thus the luminosity distance measured with GWs may differ from that measured through  EM observations. Alternatively, one can use the luminosity distance and redshift measurement of 
standard sirens to probe both cosmology and modified GW propagation. This has been demonstrated for future GW observations with advanced LIGO with its design sensitivity, Einstein Telescope (ET), and Laser Interferometer Space Antenna (LISA)~\cite{PhysRevD.98.023510,Belgacem_2019,Lagos_2019}, assuming that the luminosity distance is measured through GWs while the redshift is obtained from that of the host galaxy that is identified through  EM counterparts. If there are no associated EM counterparts, one can still use such GW sources ``dark sirens'' to probe cosmology and gravity by taking their correlation with galaxy catalogs~\cite{finke2021cosmology,10.1093/mnras/stab001}.

In this paper, we study an alternative approach of using standard sirens without EM counterparts to probe the modified GW propagation through tidal effects of BNSs. This idea was first proposed in~\cite{Messenger_2012} within the context of probing cosmology with GW observations alone. The authors in~\cite{Messenger_2012} realized that the tidal deformability that characterizes tidal effects in a BNS depends on the intrinsic mass, so together with the measurement of the redshifted mass, one can infer the source's redshift provided that one knows the nuclear matter equation of state \emph{a priori}. 

We here apply the above methodology to tests of modified GW propagation (or modified GW friction) to study how much improvement one gains from the case where one uses only BNSs with EM counterparts. We follow~\cite{PhysRevD.98.023510,Belgacem_2019} and work in a generic modified GW parametrization ($\Xi_0,n$), where $\Xi_0$ represents the ratio between the luminosity distance measured by GW and EM signals at large $z$ while $n$ denotes the redshift dependence on the ratio. Such a generic parametrization has a known mapping to theoretical constants in some specific non-GR theories~\cite{Belgacem_2019}. We carry out a Fisher analysis to derive projected bounds on $\Xi_0$ for various $n$ with ET and multi-band GW observations. The latter is a joint observation between ground- and space-based interferometers~\cite{Sesana:2016ljz,Barausse:2016eii,Isoyama_2018,Carson:2019rda,Cutler:2019krq,Carson:2019kkh,Gupta:2020lxa,Datta:2020vcj}. Here, we focus on multi-band observations between ET and DECihertz laser Interferometer Gravitational wave Observatory (DECIGO)~\cite{Kawamura_2008,kawamura2020current}.

\begin{figure}[htp]
\includegraphics[width=8.5cm]{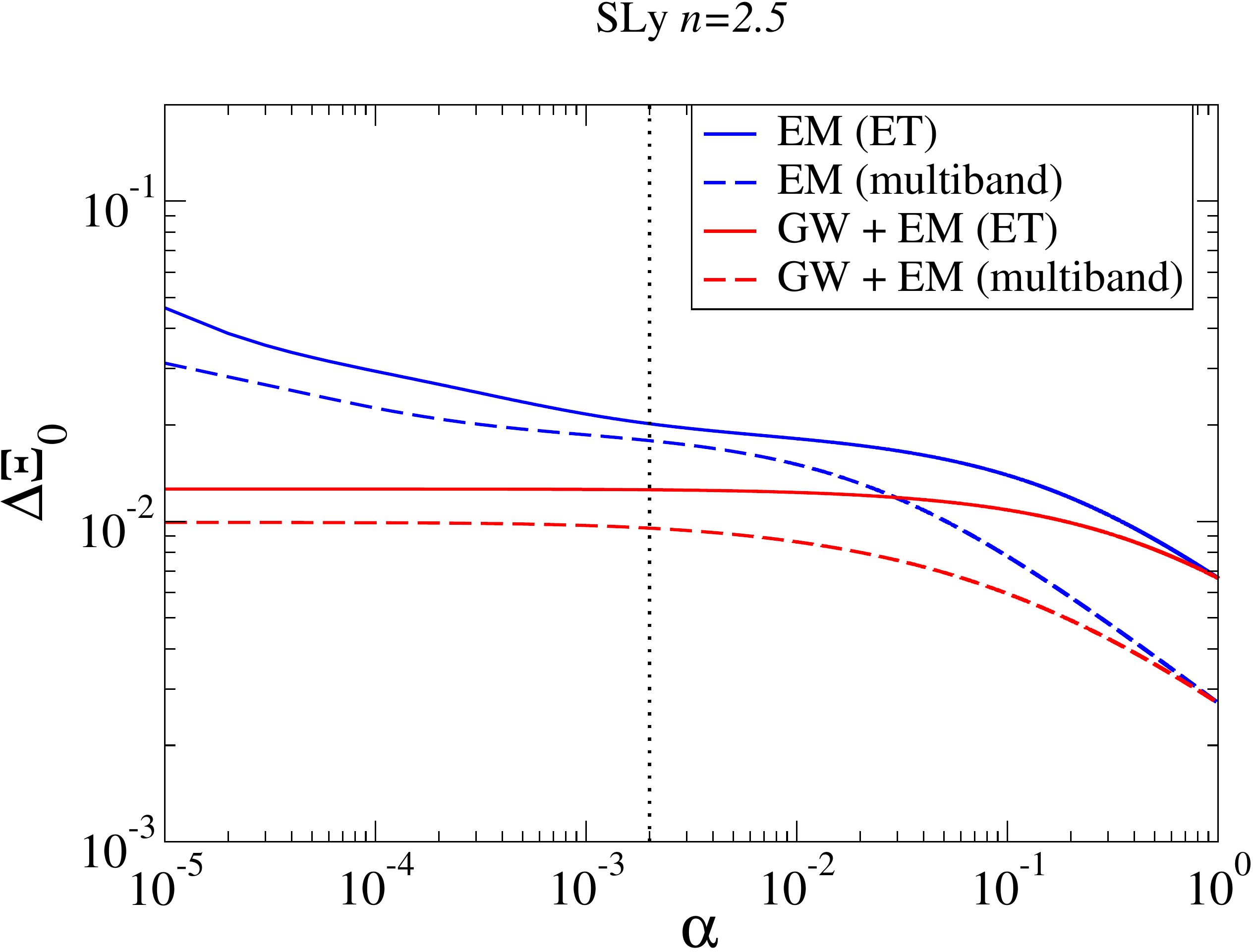}
\caption{
Measurability of the modified GW propagation parameter $\Xi_0$ as a function of the fraction $\alpha$ of the events with redshift identification through EM counterparts. We show results for (i) using only BNS events with EM counterparts (``EM'' as in Eq.~\eqref{eq:Fisher_gw+em}, as done in previous literature) and (ii) combining BNS events with and without EM counterparts (``GW+EM'' in Eq.~\eqref{eq:Fisher_combined}). We consider observations with ET alone and multi-band observations (with an observation time of 3 yrs for the latter). We use SLy EOS and set $n=2.5$ in Eq.~\eqref{eq:mod GR}. The vertical dashed line at $\alpha = 2\times 10^{-3}$ corresponds to the typical fraction of BNSs with EM counterparts~\cite{Nishizawa_2012}. Observe that the addition of BNSs without EM counterparts improves the measurability from those with EM counterpart by a factor of a few.
}
\label{fig:Result 1}
\end{figure}

We here present a brief summary of our findings. We first compute the measurability of the BNS redshift with GW observations alone, and find that multi-band observations improve the accuracy by  $\sim 50\%$ compared to the case with ET only. Next, we show the bound on the modified GW propagation parameter $\Xi_0$. Figure~\ref{fig:Result 1} presents such a bound against the fraction $\alpha$ of BNS events whose redshift is identified through EM counterparts. We choose a representative case of $n=2.5$ and SLy equation of state (EOS). Observe that the addition of BNSs without EM counterparts improves the bound by a factor of a few in the case of ET alone, and the bound further improves further if one uses multi-band observations. Although the figure is only for $n=2.5$, we find that the bound on $\Xi_0$ is insensitive to the choice of $n$. Lastly, we map the bound on $\Xi_0$ to parameters in specific non-GR theories. In the case of a scalar-tensor theory, for example, the relevant parameter can be constrained to a level of $\sim 10^{-2}$.

The organization of the rest of the paper is as follows. In Sec.~\ref{sec:level1}, we briefly introduce the formalism of how the modified luminosity distance is parameterized and the mapping between this theory-agnostic parametrization and constants in specific non-GR theories like scalar-tensor theories and phenomenological models. In Sec.~\ref{sec:2}, we will explain how to estimate uncertainties of the redshift measurement for a BNS event without EM counterpart from  the tidal effect in the gravitational waveform. Section~\ref{sec:3} describes the Fisher analysis for parameter estimation on the redshift and non-GR parameters. We present our results on the measurability of the redshift, modified GW propagation parameter and theory-specific parameters in Sec.~\ref{sec:4}. In Sec.~\ref{sec:Conclusion}, we give concluding remarks and  describe avenues for possible works.
We use the unit $G=c=\hbar=1$ throughout.

\section{\label{sec:level1}Modified Luminosity Distance}

GW sources can be used as standard sirens to probe cosmology from the relation between the luminosity distance $d_L$ and the redshift $z$~\cite{Schutz:1986gp,Abbott:2017xzu,Fishbach:2018gjp}. Such sources can also be used to probe gravity since the above relation not only depends on cosmological parameters but also on the underlying gravitational theory. 

\subsection{Formalism}

One can, in particular, probe generic theories that modifies the Hubble friction term in the propagation equation of GWs~\cite{PhysRevD.98.023510}\footnote{In general, the last term on the left hand side of Eq.~\eqref{eq:h_eq} can acquire non-GR corrections that modify the propagation speed of GWs and/or add a mass to the graviton, and an anisotropic stress source term may arise on the right hand side (see e.g.~\cite{Saltas:2014dha,Nishizawa:2017nef}).}:
\begin{equation}
\label{eq:h_eq}
\tilde h_A''+2 \mathcal{H} [1+\alpha_M(\eta)] \tilde h_A' + k_\mathrm{gw}^2 \tilde h_A = 0\,. 
\end{equation}
Here $\tilde h_A$ is the metric perturbation (or GW amplitude) in the Fourier domain with $A=+,\times$ representing the plus and cross polarization modes, a prime representing the derivative with respect to the conformal time $\eta$, $k_\mathrm{gw}$ is the wave number, $\mathcal{H} \equiv a'/a$ with $a$ denoting the scale factor, and $\alpha_M(\eta)$ is the modified friction term. The above equation reduces to the one in GR when $\alpha_M = 0$. The friction term modification affects the GW amplitude, which can be absorbed into the luminosity distance. This leads to a difference in the luminosity distance measured through GWs $d^\mathrm{gw}_L (z)$ and EM waves $d^\mathrm{em}_L (z)$ as follows~\cite{PhysRevD.98.023510,PhysRevD.102.044009}:
\begin{eqnarray}
\label{eq:mod d_L}
d^\mathrm{gw}_L (z) = d^\mathrm{em}_L (z) \mathrm{exp} \left[\int^z _0 \frac{\alpha_M (z)}{1+z} dz \right].
\end{eqnarray}

A useful parameterization has been proposed in~\cite{PhysRevD.98.023510} as
\begin{eqnarray}
\label{eq:mod GR}
\frac{d^\mathrm{gw}_L (z)}{d^\mathrm{em}_L (z)} = \Xi_0 + \frac{1-\Xi_0}{(1+z)^n}.
\end{eqnarray}
Here $\Xi_0$ corresponds to the constant ratio of the luminosity distance in the limit $z \to \infty$ while $n$ shows the redshift dependence of the ratio. GR is recovered when $\Xi_0 \to 1$ and this is the case when $z \to 0$. Such a parameterization allows us to treat the modification in the luminosity distance measurement from GWs in a generic way, and at the same time to map the modified GW propagation parameters $(\Xi_0,n)$ to theoretical constants in known gravitational theories beyond GR. 

\subsection{Mapping to Scalar-tensor Theories and Phenomenological Models}
\label{sec:mapping}

In this paper, we consider scalar-tensor theories and phenomenological models as specific examples~\cite{Belgacem_2019}. 

\subsubsection{Horndeski Theories}

Let us first review scalar-tensor theories. We consider, in particular, theories within Horndeski theories~\cite{Horndeski:1974wa}, which are most general scalar-tensor theories with field equations containing up to second order derivatives (see e.g.~\cite{Kobayashi:2019hrl} for a recent review). The action is given by~\cite{Belgacem_2019}
\begin{equation}
\label{eq:action}
S=\int d^4x\sqrt{-g}\left[\sum^5_{i=2}\mathcal{L}_i+\mathcal{L}_\mathrm{m}(g_{\mu\nu},\psi_\mathrm{m})\right]
\end{equation}
with Lagrangian densities
\begin{align*} 
\mathcal{L}_2=&G_2(\phi,X),\\
\mathcal{L}_3=&G_3(\phi,X)\Box\phi,\\
\mathcal{L}_4=&G_4(\phi,X)R-2G_{4X}(\phi,X)\left[(\Box\phi)^2-(	\nabla_\mu \nabla_\nu\phi)^2\right],\\
\mathcal{L}_5=&G_5(\phi,X)G_\mathrm{\mu\nu}\nabla^\mu \nabla^\nu\phi +\frac{1}{3}G_{5X}(\phi,X) \\
& \times \left[(\Box\phi)^3-3\Box\phi(\nabla_\mu\nabla_\nu\phi)^2+2(\nabla_\mu\nabla_\nu\phi)^3\right],
\end{align*}
where $\phi$ is the scalar field, $X\equiv \partial_\mu\phi\partial^\mu\phi$, $R$ and $G_{\mu\nu}$ represent the Ricci scalar and Einstein tensor in the Jordan frame metric $g_{\mu\nu}$. $G_i(\phi,X)$ are arbitrary functions of $\phi$ and $X$ and $G_{iX} \equiv \partial G_i/\partial X$. The matter field $\psi_\mathrm{m}$ in the Lagrangian density for matter $\mathcal{L}_m$ is minimally coupled to gravity. Given that GW170817 placed a stringent bound on the propagation speed of GWs $c_\mathrm{gw}$~\cite{TheLIGOScientific:2017qsa,Monitor:2017mdv}, we consider $G_{4X} = 0$ and $G_{5} = \mathrm{const.}$, which guarantees that $c_\mathrm{gw} = 1$~\cite{Bettoni:2016mij,Kimura:2011qn,McManus:2016kxu}. 

The correction to the Hubble friction term is related to $G_4$ through the effective Planck mass $M_\mathrm{eff}$ as
\begin{equation}
\alpha_M = \frac{d \ln M_\mathrm{eff}^2}{d\ln a}\,, \quad M_\mathrm{eff}^2 = 2 G_4\,.
\end{equation}
The modified GW propagation parameters are given by~\cite{Belgacem_2019}
\begin{equation}
\Xi_0 = \lim_{z \to \infty} \frac{M_\mathrm{eff}(0)}{M_\mathrm{eff}(z)}\,, \quad n \approx \frac{\alpha_{M0}}{2(\Xi_0 - 1)}\,,
\end{equation}
where $\alpha_{M0}$ is $\alpha_M$ at the present time.

As an example of Horndeski theories, we consider $f(R)$ gravity where the Einstein-Hilbert action is modified with $R \to R + f(R)$ for an arbitrary function $f$. $G_4$ then becomes 
\begin{equation}
G_4 = \frac{1+f_R}{2} M_\mathrm{P}^2, 
\end{equation}
where $M_\mathrm{P}$ is the Planck mass that is related to the effective Planck mass as $M_\mathrm{P} = \lim_{z \to \infty} M_\mathrm{eff}(z) $. $f_R \equiv f'(R)$ and a prime represents a derivative with respect to $R$. For such a model, $\Xi_0$ and $n$ are given by
\begin{eqnarray}
\Xi_0 &=& \sqrt{1+f_{R0}} \approx 1 + \frac{1}{2}f_{R0}, \\
n & \approx &\left( \frac{f'_R}{f_R} \right)_0,
\end{eqnarray}
where the subscript 0 corresponds to the present value. 
In particular, we consider a model proposed by Hu and Sawicki (HS). $\Xi_0$ and $n$ for the HS $f(R)$ gravity are given in Table~\ref{tab:mapping} where $\bar n$ is a positive integer and $\Omega_M$ is the matter energy density parameter.  

$f(R)$ gravity is a special case of Brans-Dicke theory~\cite{PhysRev.124.925}. $G_4$ and $G_2$ in the latter theory is given by 
\begin{equation}
    G_4(\phi)\equiv \frac{M_\mathrm{P}^2\,\phi}{2}\,,
\end{equation}
and $G_2 = -U(\phi) + X \omega(\phi)/\phi$, where $\omega$ is the Brans-Dicke function and $U$ is the scalar field potential. The theory reduces to $f(R)$ gravity when $\omega = 0$. The mapping of $(\Xi_0,n)$ to Brans-Dicke theory is given in Table~\ref{tab:mapping}, where $\delta \phi_0 \equiv  \phi_0 -1$.

\renewcommand{\arraystretch}{1.5}
\begin{table} [htp]
\begin{tabular}{c c c}
\hline
\hline
Non-GR Model & $\Xi_0-1$&$n$ \\
\noalign{\smallskip}
\hline
\hline
\noalign{\smallskip}
HS $f(\mathrm{R})$~\cite{PhysRevD.76.064004} & $\frac{1}{2}f_\mathrm{R0}$& $\frac{3(\Bar{n}+1)\Omega_M}{4-3\Omega_M}$ \\
\hline
designer  $f(\mathrm{R})$~\cite{Song_2007}&$-0.24\Omega_M^{0.76} B_0$ & $3.1\Omega_M^{0.24}$ \\
\hline
Brans-Dicke~\cite{PhysRev.124.925}&$\frac{1}{2} \delta \phi_0$&$\frac{3(\Bar{n}+1)\Omega_M}{4-3\Omega_M}$\\
\hline\hline
power law $\alpha_M$~\cite{Bellini_2014}
&$\frac{\alpha_{M0}}{2\Bar{n}}$ &$\Bar{n}$\\
\hline
DE density $\alpha_M$~\cite{Bellini_2014,Simpson_2012}
&$-\frac{\alpha_{M0}}{6\Omega_\Lambda \mathrm{ln}\Omega_M}$&$-\frac{3\Omega_\Lambda}{\mathrm{ln} \Omega_M}$\\
\hline
power law $M_\mathrm{eff}$~\cite{Lombriser_2016}
&$\frac{1}{2}\Omega_+$&$\Bar{n}$\\
\hline
\hline
\end{tabular}
\caption{\label{tab:mapping} 
Mapping of the modified GW propagation parameters $(\Xi_0,n)$ to parameters in scalar-tensor theories (top) and phenomenological models for $\alpha_M$ or the effective Planck mass  $M_{\mathrm{eff}}$  (bottom)~\cite{Belgacem_2019}.
}
\end{table}

\subsubsection{Phenomenological Models}
\label{sec:phenom_alphaM}

The second model we consider is a phenomenological parameterization on $\alpha_M$ motivated by a time-varying effective Planck mass $M_\mathrm{eff}$. 
We consider two different parameterization for $\alpha_M$: 
\begin{itemize}
\item [(i)] power law: 
\begin{equation}
\label{eq:alphaM_powerlaw}
\alpha_M=\alpha_{M0} a^{\Bar{n}};
\end{equation}
\item [(ii)] dark energy density:
\begin{equation}
\label{eq:alphaM_DE}
\alpha_M=\alpha_{M0}\frac{\Omega_\Lambda(a)}{\Omega_{\Lambda 0}},
\end{equation}
where $\Omega_{\Lambda}$ is the dark energy density parameter.

\end{itemize}
Once again, $\Xi_0$ and $n$ for these models are summarized in Table~\ref{tab:mapping}.

In Appendix~\ref{sec:Appendix}, we review other models within Horndeski theories and phenomenological classes, and give the mapping to $(\Xi_0,n)$ in Table~\ref{tab:mapping}.

\section{\label{sec:2}Redshift Inference through Tidal Effects}

To probe gravity from the luminosity distance-redshift relation in Eq.~\eqref{eq:mod GR}, one needs an independent measurement of $d_L$ and $z$. The former is measured from the amplitude of GWs while the latter is more challenging to measure as it typically degenerates with the mass. If a BNS event has an associated EM counterpart, one can use the redshift information of the host galaxy, which has been used for GW170817 to measure the Hubble constant~\cite{Abbott:2017xzu} and also to give future forecasts on testing the modified GW propagation~\cite{PhysRevD.98.023510}. 
However, BNS events with EM counterparts are expected to be rare, with a fraction of only $\sim 10^{-3}$ or so~\cite{Nishizawa_2012}.

An alternative method to measure the redshift with GW observations alone is to use the tidal effect~\cite{Messenger_2012}. Such an effect in BNS is characterized by tidal deformabilities or Love numbers that depend on the intrinsic (source-frame) masses of NSs. Together with the redshifted mass measurement, one can break the degeneracy between the redshift and the mass to extract the former. This method requires one to know the nuclear matter equation of state \emph{a priori} which still has relatively large uncertainties. One may use future GW observations of nearby BNS sources ($z \lesssim 0.1$) with EM counterparts to determine the equation of state, and use those of BNSs with large $z$ to probe the modified GW propagation~\cite{Messenger_2012,Wang_2020}. 

Let us explain this tidal method in more detail by taking NRTidalv2~\cite{Dietrich_2019,PhysRevD.96.121501} as an example. The tidal contribution to the gravitational wave phase in the frequency domain is given by 
\begin{eqnarray}
\label{eq:tidal}
	\psi_\mathrm{T} (x) = -\frac{13}{8 \nu} \kappa_\mathrm{eff}  x^{5/2} P (x),
\end{eqnarray}
where  $x = (\pi M_z f)^{1/3}$ with $M_z = (1+z)M$ is the total redshifted mass with $M$ representing the intrinsic total mass, $\nu$ representing the symmetric mass ratio $m_A m_B/(m_A+m_B)^2$ and $f$ is the observed GW frequency. 
$P(x)$ is a Pad\'e-resummed function given by
\begin{eqnarray}
	P(x) = \frac{1+ n_1 x +n_{3/2} x^{3/2} +n_2 x^2+n_{5/2} x^{5/2}+n_3 x^3}{1+d_1 x+d_{3/2} x^{3/2}+d_2 x^2}, \nonumber \\
\end{eqnarray}
where the coefficients can be found in~\cite{Dietrich_2019}. $\kappa_\mathrm{eff}$ is related to the tidal Love number $k$ as
\begin{eqnarray}
	\kappa_\mathrm{eff} = \frac{2}{13} \left[ \left(1+12\frac{X_B}{X_A}\right)\left(\frac{X_A}{C_A} \right)^5 k^A + (A \leftrightarrow B)\right].
\end{eqnarray}
Here, subscript $A$ and $B$ denotes the two component stars, $X_{A} \equiv m_{A}/M$ and the compactness is given by $C_{A} \equiv m_{A}/R_{A}$ with the stellar radius $R_A$.
Since $k$ depends on the intrinsic stellar mass instead of the redshifted one, the tidal effect can be used to extract the redshift information from a GW observation alone.

\section{\label{sec:3}Fisher Analysis}

In this paper, we carry out a parameter estimation based on a Fisher analysis~\cite{Cutler:1994ys}, which is valid for sources with sufficiently large signal-to-noise ratios (SNRs). We perform two different Fisher calculations, one for the redshift estimate and another for the modified GW propagation parameter estimate. Below, we will explain each of these Fisher analyses in turn.

\subsection{Redshift Estimate}

The first step is to estimate the measurability of the redshift by using template gravitational waveforms of BNSs, which we take as the (non-spinning) IMRPhenomD-NRTidalv2 waveform~\cite{Dietrich_2019,PhysRevD.96.121501,PhysRevD.93.044006,PhysRevD.93.044007}. It consists of the IMRPhenomD waveform for point-particle binaries with an updated tidal effect added to the phase. The waveform $\tilde h$ in the frequency domain can be written as  
\begin{eqnarray}
	\label{eq:waveform in f}
	\tilde{h}(f) = \tilde{A}(f) e^{-i\psi(f)},
\end{eqnarray}
where $\tilde A$ is the IMRPhenomD amplitude\footnote{The NRTidal waveform also has a tidal correction to the amplitude, though 
the tidal effect is mostly determined from the phase and thus we do not include such effects in the amplitude for simplicity.} while $\psi$ is the phase given by\footnote{In this paper, we include the tidal phase only in the inspiral part of the IMRPhenomD phase, though we have checked that our results are unaffected even if we include the tidal phase also in the intermediate portion of the waveform.}
\begin{eqnarray}
	\label{eq:phase in f}
	\psi(f) = \psi_\mathrm{pp}(f)+\psi_\mathrm{T}(f).
\end{eqnarray}
Here $\psi_\mathrm{pp}$ is the (non-spinning) point-particle term that is taken from the IMRPhenomD waveform while $\psi_\mathrm{T}$ is the tidal contribution given in Eq.~\eqref{eq:tidal} that is parameterized by the Love number $k$. 
In our analysis, we use the tidal deformability $\lambda \equiv (2/3) R^5 k$, which is a function of the NS mass $m$. It is convenient to Taylor expand $\lambda(m)$ about a fiducial mass $m_0$ as~\cite{Messenger_2012,Wang_2020}
\begin{eqnarray}
\lambda &=& \lambda_0 + \lambda_1 (m-m_0) + \mathcal{O}[(m-m_0)^2] \nonumber \\
&=& \tilde{\lambda}_0 + \tilde{\lambda}_1 m+ \mathcal{O}[(m-m_0)^2], 
\end{eqnarray}
where $\lambda_i$ are the Taylor coefficients about $m_0$ while $\tilde \lambda_0 = \lambda_0 - \lambda_1 m_0$ and $\tilde \lambda_1 = \lambda_1$.

One can compute the measurability of parameters $\theta^i$ from a Fisher matrix as follows. We first assume that the detector noise is stationary and Gaussian. Then, the probability distribution of $\theta^i$ becomes also Gaussian as 
\begin{eqnarray}
	\label{eq:Fisher 1}
	p(\theta^i) \propto 
    \mathrm{exp} \left[-\frac{1}{2} \Gamma_{ij} \left(\theta^i - \hat \theta^i\right)  \left(\theta^j - \hat \theta^j\right) \right],
\end{eqnarray}
where $\hat \theta^i$ are the maximum likelihood parameters. 
$\Gamma_{ij}$ is the Fisher matrix defined as
\begin{equation}
\label{eq:Fisher}
\Gamma_{ij} = 4\Re \int^{f_\mathrm{high}}_{f_\mathrm{low}} \frac{\partial_i \tilde h \partial_j \tilde h}{S_n(f)} df,
\end{equation}
where $\partial_i \equiv \partial/\partial \theta^i$ while $S_n$ is the noise spectral density. $f_\mathrm{high}$ and $f_\mathrm{low}$ are the high and low frequency cutoffs to be discussed later. 
\begin{equation}
\tilde{\Gamma}_{ij} = \sum_{A} \Gamma_{ij}^{(A)}, 
\end{equation}
where $A$ is the label of each detector. 
Finally, the $1\sigma$ root-mean-square error on $\theta_i$ is given by
\begin{eqnarray}
    \label{eq:Fisher uncertainty}
    \Delta \theta^i = \sqrt{(\tilde{\Gamma}^{-1})_{ii}}.
\end{eqnarray}

\begin{figure}[t]
\includegraphics[width=8.5cm]{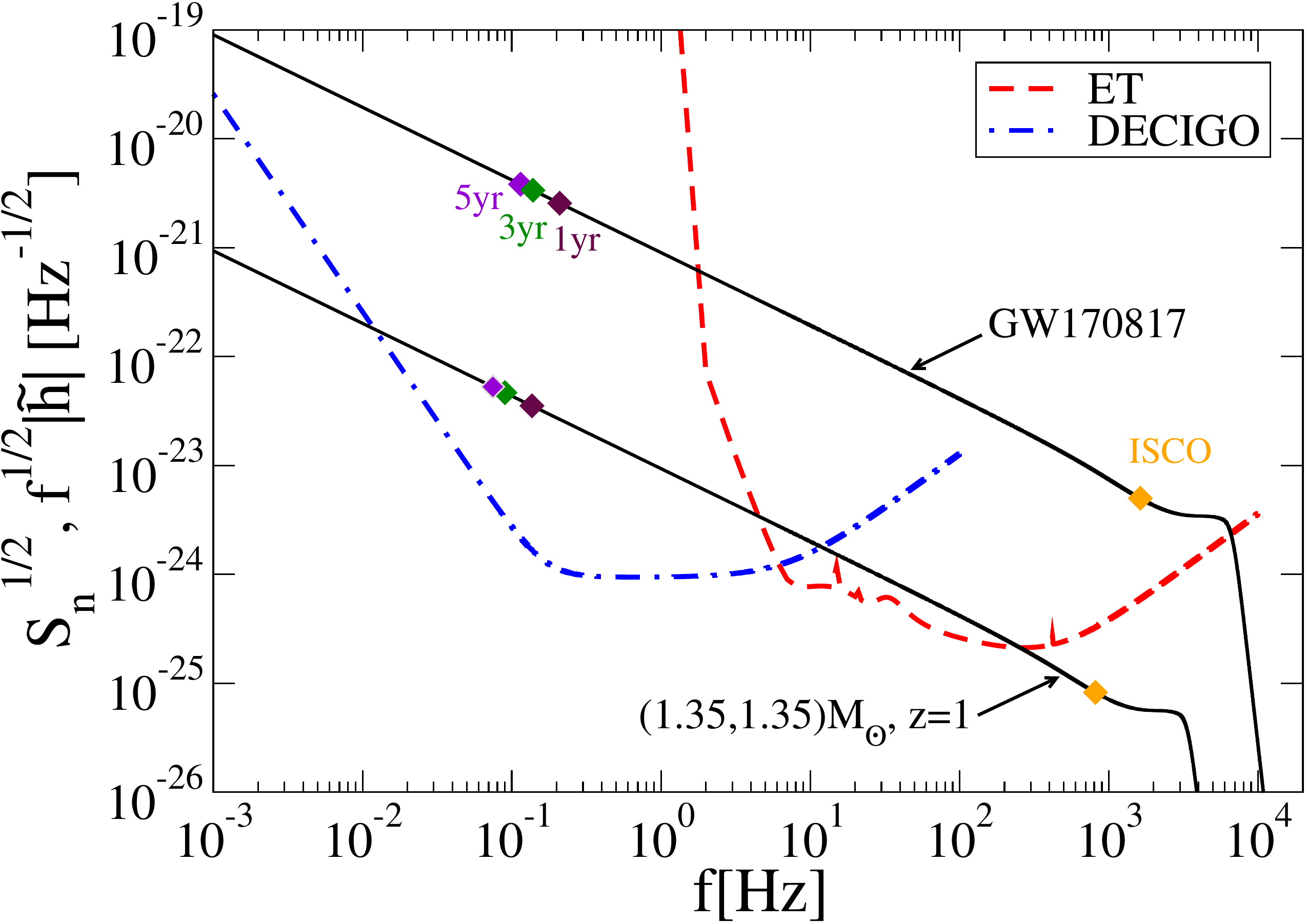}
\caption{The noise spectral densities for ET~\cite{Hild_2011} and DECIGO~\cite{Yagi_2011}. We also present the GW spectrum for GW170817 and a BNS with $(1.35,1.35)M_\odot$ at $z=1$. For each GW spectrum, we show the frequency at ISCO and that at 1yr, 3yr and 5yr before coalescence.}
\label{fig:NoiseCurve}
\end{figure}

In Fig.~\ref{fig:NoiseCurve}, we present $S_n$ for ET and DECIGO, together with the GW spectrum for GW170817 and a BNS with $(1.35,1.35)M_\odot$ at $z=1$. 
For ET, we choose the low and high frequency cutoffs in the Fisher matrix in Eq.~\eqref{eq:Fisher} as
\begin{equation}
f_\mathrm{low}^{\ET} = 1\mathrm{Hz}, \quad f_\mathrm{high}^{\ET} = \min(f_\ISCO,f_\mathrm{cont}),
\end{equation}
where 
\begin{equation}
 \label{eq: isco}
 f_\ISCO = \frac{1}{6^{3/2} \pi M_z},
 \end{equation}
 is the frequency at the innermost stable circular orbit (ISCO) while $f_\mathrm{cont}$ is the (redshifted) contact frequency of two NSs and is given by
 \begin{equation}
 \label{eq: touch f}
 f_\mathrm{cont} =  \frac{1}{2^{3/2}\pi (1+z)} \sqrt{\frac{M}{R^3}},
 \end{equation}
 for an equal-mass BNS with $R$ representing the stellar radius. For a NS with a soft (stiff) EOS, the radius is relatively small (large), and $f_\ISCO < f_\mathrm{cont}$ ($f_\ISCO > f_\mathrm{cont}$). On the other hand, for DECIGO, we choose the low and high cutoff frequencies as 
 \begin{eqnarray}
	\label{eq:ET range}
	f_\mathrm{low}^\DECIGO& = & 0.233 \left(\frac{1 M_\odot}{\mathcal{M}_z}\right)^{5/8} \left(\frac{1 \mathrm{yr}}{T_\mathrm{obs}}\right)^{3/8} \mathrm{Hz}, \nonumber \\
    f_\mathrm{high}^\DECIGO &=&100 \mathrm{Hz},
\end{eqnarray}
 where  $\mathcal{M}_z = M_z \eta^{3/5}$ is the redshifted chirp mass,  $T_\mathrm{obs}$ is the observation time and $f_\mathrm{low}^\DECIGO$ corresponds to the (redshifted) frequency at $T_\mathrm{obs}$ before coalescence.

Let us now explain parameters $\theta^i$ specific to our analysis. We use the sky-averaged waveform and the parameters are given by
\begin{eqnarray}
\label{eq:first Fisher}
	\theta^i = \left(\mathrm{ln}\mathcal{M}_z, \eta,t_c,\phi_c,\mathrm{ln}A,\mathrm{ln}z \right).
\end{eqnarray}
Here $\eta = m_1 m_2/M^2$ is the symmetric mass ratio with individual masses $m_A$, and $t_c$ and $\phi_c$ are the coalescence time and phase respectively. The amplitude parameter $A$ is given by $A =  \mathcal{M}_z^{5/6}/(\sqrt{30} \pi^{2/3} d_L^{\mathrm{gw}})$, which corresponds to the leading, sky-averaged amplitude in the frequency domain without the frequency dependence~\cite{Berti_2005}. We assume the tidal parameters $\tilde{\lambda}_0$ and $\tilde{\lambda}_1$ are known \emph{a priori} from BNSs with $z < 0.1$ (we discuss how the imperfect knowledge of the EOS affects the measurability of the redshift in Appendix~\ref{sec:Appendix1}).
Regarding fiducial values for Fisher analyses, we choose $m_1 = m_2 = m_0 = 1.35M_\odot$, $t_c =0$, $\phi_c=0$, and vary $z$ or $d_L^\mathrm{gw}$. Fiducial values for $\tilde \lambda_0$ and $\tilde \lambda_1$are summarized in Table~\ref{tab:lambdapriors} in Appendix~\ref{sec:Appendix1} for three EOSs as representatives of soft, intermediate and stiff classes: SLy~\cite{refId0}, MPA1~\cite{MUTHER1987469} and MS1~\cite{MULLER1996508}.

\subsection{
Parameter Estimation for Modified GW Propagation }

We now move onto explaining the second Fisher analysis for estimating the measurability of cosmological parameters and the modified GW propagation parameter. We consider a spatially-flat Universe and work on the following four parameters~\cite{PhysRevD.98.023510}: 
\begin{eqnarray}
\label{eq:parameter2}
	p^i = \left(\mathrm{ln}H_0, \mathrm{ln}\Omega_M, w_0, \Xi_0 \right).
\end{eqnarray}
Here $H_0$ is the Hubble constant, $\Omega_M$ is the matter energy density parameter at present time, $w_0$ is the equation of state parameter for dark energy~\cite{doi:10.1142/S0218271801000822,PhysRevLett.90.091301}\footnote{The equation of state for dark energy is given by $P_\DE = w_0 \varepsilon_\DE $ where $P_\DE$ and $\varepsilon_\DE$ are the pressure and energy density of dark energy.} while $\Xi_0$ is the modified GW propagation parameter in Eq.~\eqref{eq:mod GR}. The luminosity distance measured by EM observations depends only on the first three parameters in Eq.~\eqref{eq:parameter2} as
\begin{eqnarray}
	\label{eq:d_l}
	d_L^\mathrm{em}(z) = (1+z) \int^z_0 \frac{\mathrm{d}\tilde{z}}{H(\tilde z)},
\end{eqnarray}
with the Hubble parameter given by
\begin{eqnarray}
	\label{eq:Hz}
	H(z) = H_0 \sqrt{\Omega_M (1+z)^3+(1-\Omega_M)(1+z)^{3 (1+\omega_0)}}. \nonumber \\
\end{eqnarray}

We can construct a Fisher matrix to estimate the measurability of the parameters $p^i$ by studying how $\ln d_L^\mathrm{gw}$ depends on each of these parameters and comparing it with a measurement error on $\ln d_L^\mathrm{gw}$. Combining information from multiple events, we can write down the Fisher matrix as~\cite{Wang_2020}
\begin{equation}
	\label{eq:individual Fisher}
	F_{ij}^\mathrm{(A)} = \sum_a \frac{(\partial \mathrm{ln}d_L^\mathrm{gw}/\partial p_i)(\partial \mathrm{ln}d_L^\mathrm{gw} /\partial p_j)}{(\Delta \ln d_L ^\mathrm{gw} )^2_{\mathrm{(A)}}}\Bigg|_{a}.
\end{equation}
Here $a$ labels each BNS 
while $A = (\mathrm{gw},\mathrm{em})$ labels whether the redshift is measured from GWs through the tidal effects or from EM counterparts. $(\Delta \ln d_L ^\mathrm{gw} )^2_{\mathrm{(A)}}$ is the total error on $\ln d_L^\mathrm{gw}$ given by
\begin{eqnarray}
\label{eq:dL_tot}
(\Delta \ln d_L ^\mathrm{gw} )^2_{\mathrm{(A)}} &=& (\Delta \ln d_L^\mathrm{gw})_\mathrm{gw}^2 + \epsilon_A \left(  \frac{\partial \ln d_L^\mathrm{gw}}{\partial z} \Delta z
\right)^2  \nonumber \\
&&  + (\Delta \ln d_L^\mathrm{gw})_\mathrm{lens}^2\,,
\end{eqnarray}
with $\epsilon_\mathrm{gw} = 1$ and $\epsilon_\mathrm{em} = 0$.
The first term on the right hand side is the measurement error on $\ln d_L^\mathrm{gw}$ through GWs, the second term is due to the measurement error on the redshift, while the last term is due to the gravitational lensing given by~\cite{Sathyaprakash_2010}
\begin{equation}
	(\Delta \ln d_L^\mathrm{gw})_\mathrm{lens} \simeq 0.05 z .
\end{equation}
The first two terms are computed from $\Gamma_{ij}$ in the previous subsection, either with ET alone or with the multi-band observations. For BNSs with redshift identified from EM counterparts, the measurement error on the redshift is typically negligible and we drop the second term in Eq.~\eqref{eq:dL_tot} (i.e. $\epsilon_\mathrm{em}=0$) for such cases.

\begin{figure}[t]
\includegraphics[width=8.5cm]{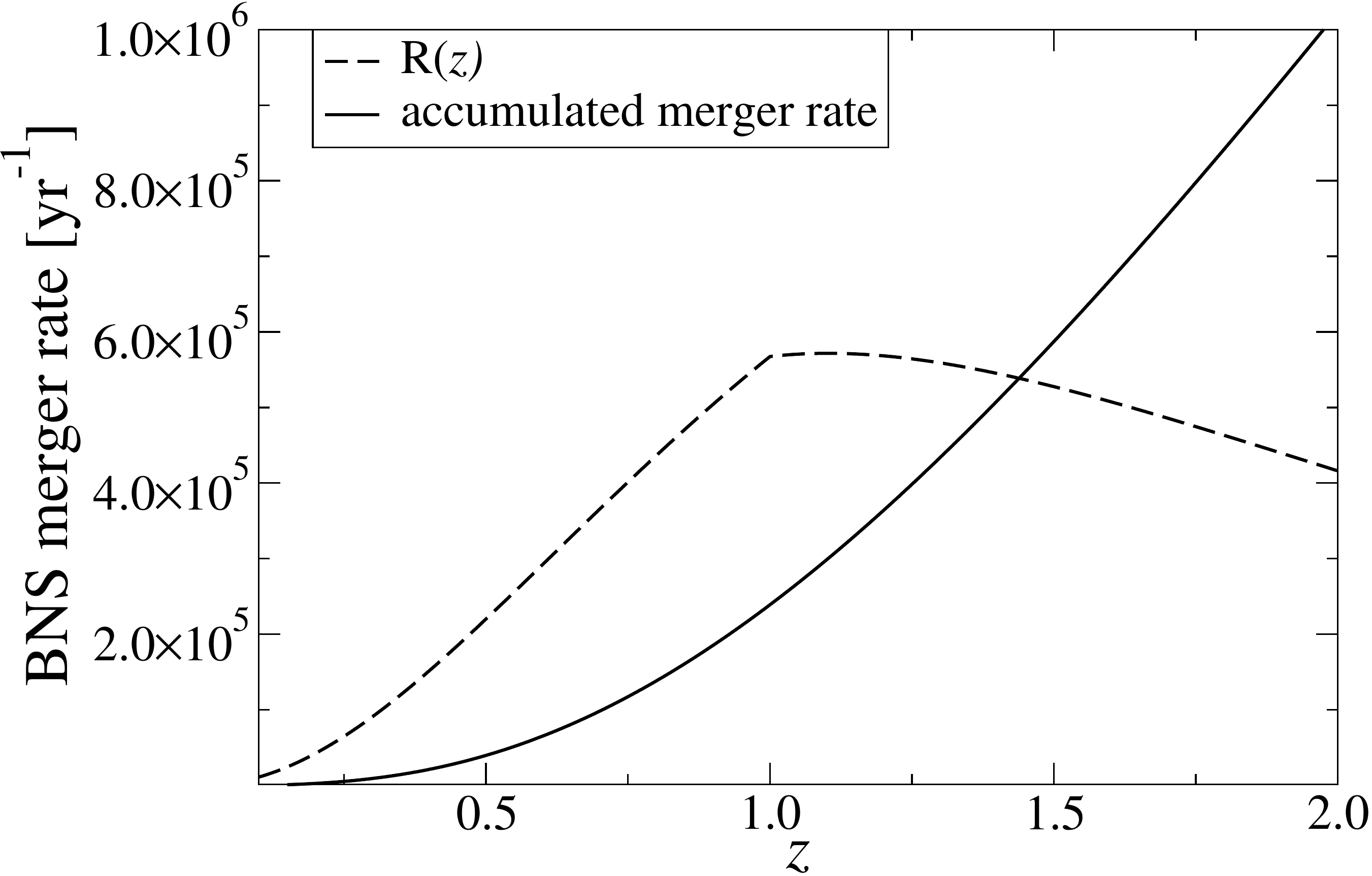}
\caption{The BNS merger rate per unit redshift ($\mathcal{R}(z)$) and the accumulated number of events up to a given redshift ($\int_0^z \mathcal{R}(z) dz$) as a function of $z$.} 
\label{fig:R(z)}
\end{figure}

In this paper, we follow~\cite{Wang_2020} and assume that all BNSs are identical except for their redshifts. Under this assumption, one can turn the summation in $F_{ij}$ into an integral as 
\begin{equation}
  \label{eq:int fisher}
   F_{ij}^\mathrm{(A)} = \int^{z_{\max}}_{z_{\min}} \frac{(\partial \mathrm{ln}d_L^\mathrm{gw} /\partial p_i)(\partial \mathrm{ln}d_L^\mathrm{gw} /\partial p_j)}{(\Delta \ln d_L ^\mathrm{gw} )^2_{\mathrm{(A)}}} \mathcal{R}(z) \mathrm{d}z .
\end{equation}
We choose the minimum and maximum redshifts ($z_{\min}$ and $z_{\max}$) as $z_{\min} = 0.1$ and $z_{\max} = 2$. This is because we use BNS sources with $z < 0.1$ (with EM counterparts) to determine the NS EOS while the SNR becomes too small for detection when $z > 2$~\footnote{The SNR for a sky-averaged BNS at $z=2$ with ET is 4.5 which may be smaller than the detection threshold SNR, though the latter may be reduced if we have additional information from DECIGO for multi-band observations (see e.g.~\cite{Wong_2018} for a related work).}. $\mathcal{R}(z)$ is the distribution of BNS mergers which is given by~\cite{Cutler_2006} 
\begin{equation}
	\label{eq:f(z)}
	\mathcal{R}(z) = \frac{4 \pi r^2(z) \dot{n}_0s(z)}{H(z) (1+z)} T_\mathrm{obs},
\end{equation}
in which $\dot{n}_0 = 10^{-6}\mathrm{Mpc}^{-3} \mathrm{yr}^{-1}$~\cite{PhysRevX.9.031040} is the current BNS merger rate, $r(z)$ is the comoving distance, and
\begin{eqnarray}
	s(z) = \begin{cases} 1+2z & (z \leq 1) \\ \frac{3}{4} (5-z) &(1<z\leq5) \end{cases},
\end{eqnarray}
shows the redshift evolution of the merger rate. We show the BNS merger rate  within  each  redshift bin  and  the accumulated  merger rate  up  to  a given redshift in Fig.~\ref{fig:R(z)}.

Given that various cosmological observations, including cosmic microwave background (CMB), baryon acoustic oscillation (BAO) and supernovae, measured cosmological parameters with some errors, one can impose prior on such parameters for our Fisher analysis. For simplicity, we impose Gaussian priors with standard deviation $\sigma^0_{p^i}$ for each parameter. The Fisher matrix for BNSs with redshift identification due to EM counterparts is given by 
\begin{equation}
\label{eq:Fisher_gw+em}
\tilde{F}_{ij}^\mathrm{(em)} = \alpha F_{ij}^\mathrm{(em)} + \frac{\delta_{ij}}{(\sigma^0_{p^i})^2}\,.
\end{equation}
Here $\alpha$ is the fraction of total BNSs with which the redshifts are identified through their EM counterparts~\cite{Nishizawa_2012}.
One can further add BNSs whose redshift is identified through the tidal measurement of GWs as
\begin{equation}
\label{eq:Fisher_combined}
\tilde{F}_{ij}^\mathrm{(gw+em)} =(1-\alpha) F_{ij}^\mathrm{(gw)} +  \alpha F_{ij}^\mathrm{(em)} + \frac{\delta_{ij}}{(\sigma^0_{p^i})^2}\,.
\end{equation}
The 1-$\sigma$ root-mean-square error on $p^i$ can be estimated as
\begin{equation}
\Delta p^i = \sqrt{(\tilde F^{-1})_{ii}}\,.
\end{equation}

We end this section by describing the fiducial values and priors for $p^i$. For the former, we use $H_0 = 67.64 \, \mathrm{km~s}^{-1}\mathrm{Mpc}^{-1}, \Omega_M = 0.3087, w_0 = -1, \Xi_0 = 1$. This corresponds to the $\Lambda$CDM model in GR with the first two parameter values being the best-fit values from CMB, BAO and supernovae observations~\cite{PhysRevD.98.023510}. For the prior, we use~\cite{PhysRevD.98.023510} 
\begin{equation}
\label{eq:CMB+BAO+SNe}
 (\sigma^0_{\Xi_0}, \sigma^0_{\omega_0}, \sigma^0_{\ln H_0},\sigma^0_{\ln\Omega_M}) = (\infty, 0.0535, 0.018, 0.039),
\end{equation}
which is obtained from the same datasets as those for the above fiducial values.

\section{\label{sec:4}Results}

\begin{figure}[t]
\includegraphics[width=8.5cm]{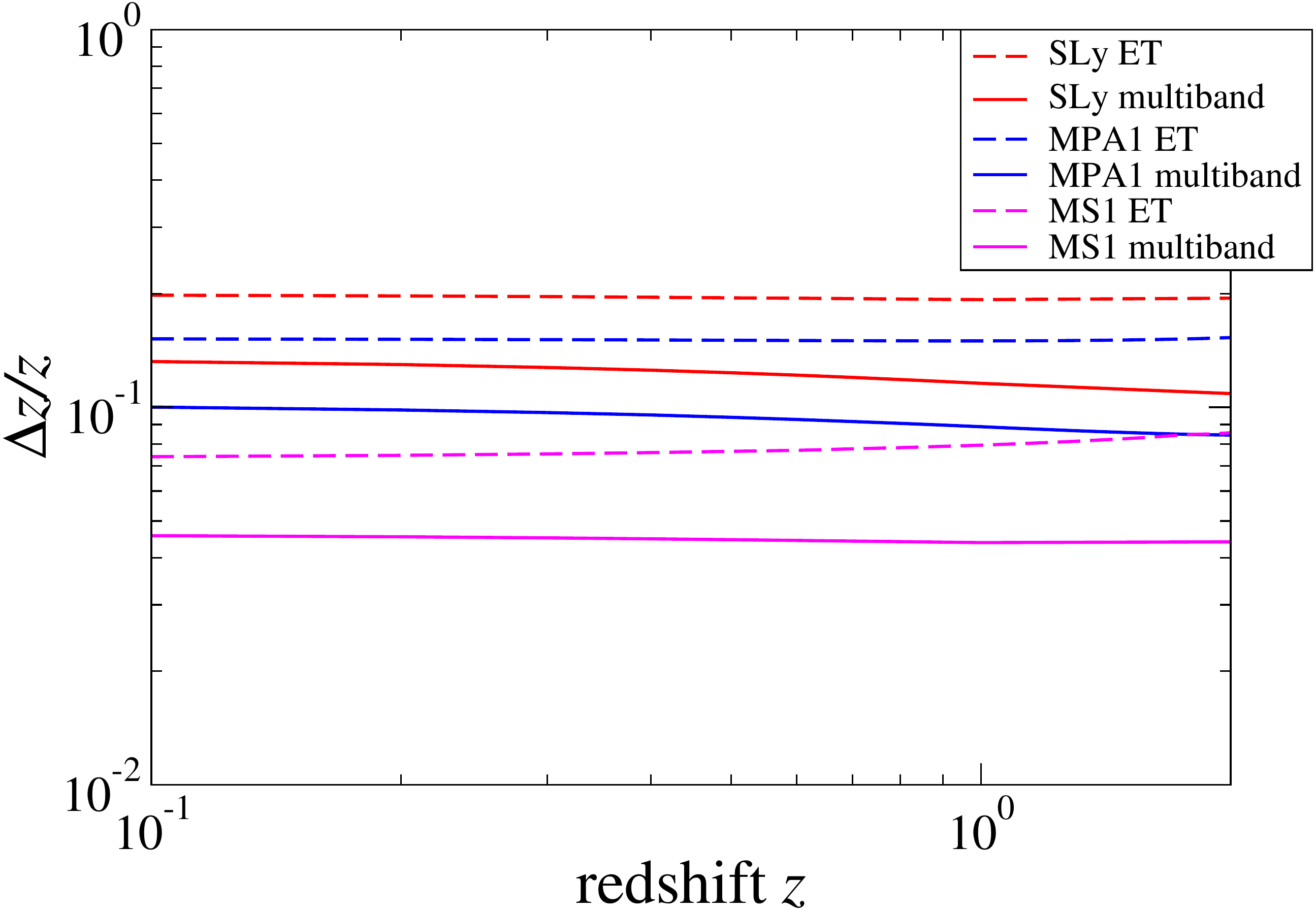}
\caption{The fractional uncertainty of the redshift as a function of the source redshift with ET and multi-band GW observations (with a 3-yr observation for the latter). We present the results for SLy (soft)~\cite{refId0}, MPA1 (intermediate)~\cite{MUTHER1987469} and MS1 (stiff)~\cite{MULLER1996508} EOSs. Notice that the uncertainty is insensitive to $z$ and is larger for softer EOSs.
}
\label{fig:Redshift}
\end{figure}

We now present our main results. We first show the measurability of redshift with GW observations. We next use this to compute the measurability of the modified GW propagation parameter $\Xi_0$ and cosmological parameters. We finally map the projected bounds on $\Xi_0$ to those on example theories within the Horndeski class and example phenomenological models.

\subsection{Redshift Inference}
\label{sec:redshift_inference}

We begin by showing the measurement accuracy of $z$ with GW observations using ET and multi-band (ET + DECIGO) detections in Fig.~\ref{fig:Redshift} for the three representative EOSs. Observe that the redshift can be measured to $\mathcal{O}(10\%)$ and is insensitive to the BNS redshift. Notice also that the measurability of $z$ increases as the EOS becomes stiffer. This is because the NS radius becomes larger and the tidal effect in turn becomes stronger. We further see that the multi-band detection improves the measurability of $z$ from the case with ET alone by $\sim 50\%$. The result for ET in Fig.~\ref{fig:Redshift} is consistent with that in~\cite{Messenger_2012}. The difference originates from using e.g. different point-particle waveforms (IMRPhenomD v.s. Taylor F2) and tidal effects (5 and 6PN v.s. NRTidal fit).

\begin{figure}[htp]
\includegraphics[width=8.5cm]{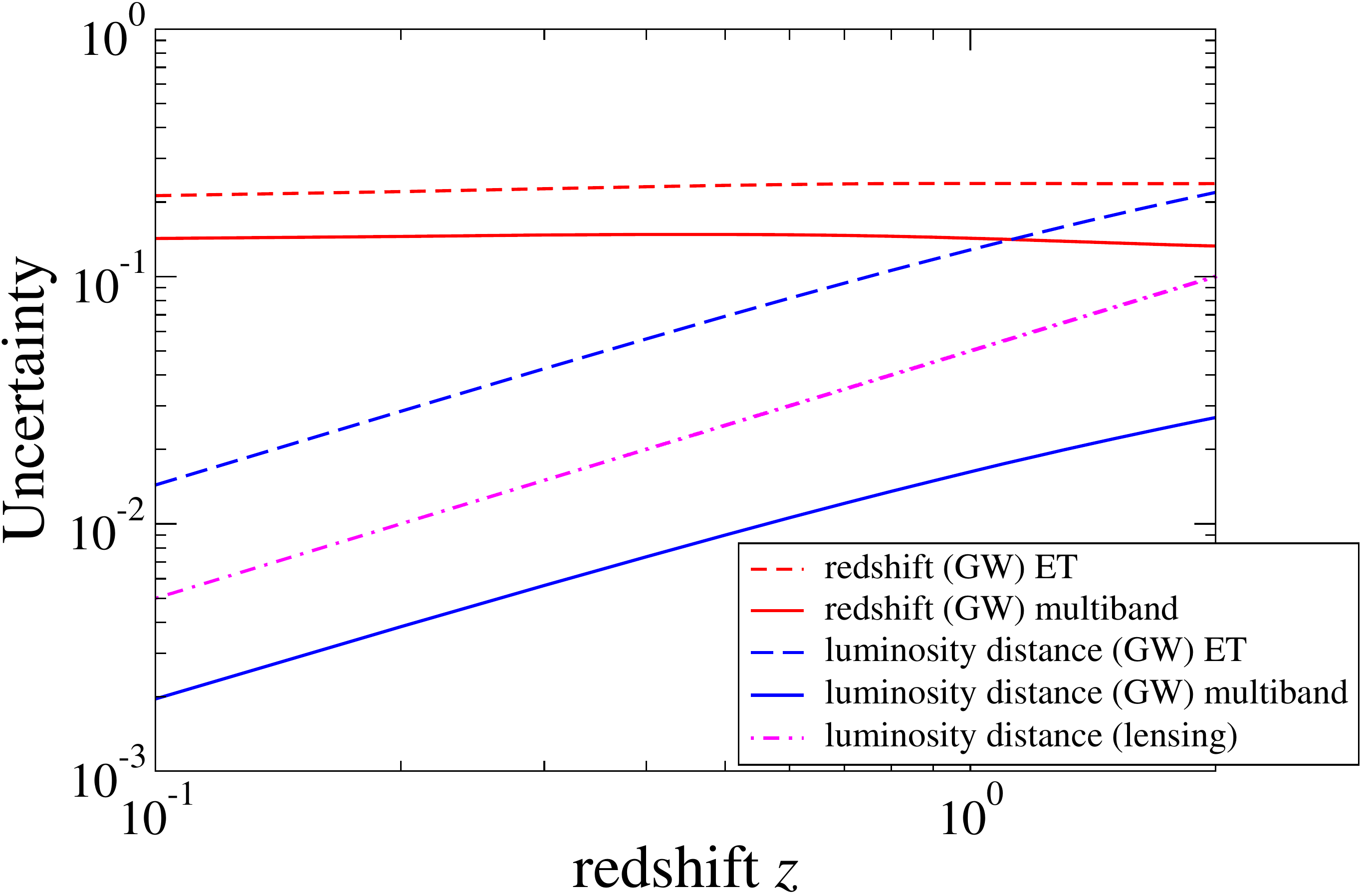}
\caption{Various sources of the luminosity distance measurement uncertainties in Eq.~\eqref{eq:dL_tot}  as a function of the BNS redshift. We use the SLy EOS and assume multi-band observations (with a 3-yr observation time). Notice that the redshift uncertainty dominates the error budget throughout.}
\label{fig:ErrorCompare}
\end{figure}

Before showing bounds on $\Xi_0$, let us first present in Fig.~\ref{fig:ErrorCompare}  different errors on the luminosity distance (Eq.~\eqref{eq:dL_tot}) in the second Fisher matrix $F_{ij}$. We chose SLy EOS and the multi-band observation. Notice that the error propagated from the redshift measurement in Fig.~\ref{fig:Redshift} dominates the other two errors (direct measurement of $d_L^\mathrm{gw}$ from GWs and the lensing) for both ET alone and multi-band observations. On the other hand, when there is an EM counterpart, the error from redshift is negligible and it is the lensing (direct luminosity distance measurement) error that gives the dominant contribution for multi-band (ET alone) observations.

\subsection{Constraints on GW propagation parameter $\Xi_0$}
\label{sec:Xi_0}

Having the redshift measurability at hand, we next present the measurability of the modified GW propagation parameter $\Xi_0$. Figure~\ref{fig:Result 1} in Sec.~\ref{sec:intro} presents such a measurement error on $\Xi_0$ for $n=2.5$ against the fraction $\alpha$ of the redshift identification of BNSs through EM counterparts for ET and multi-band observations. We show the results using BNSs with EM counterparts only (whose redshifts are identified), and combining BNSs with and without the counterparts. We chose the SLy EOS and an observation time of 3 yrs for DECIGO in the multi-band observations (see Appendix~\ref{sec:Appendix2} for how the results change with a different choice of EOSs and observation time).  Notice first that the addition of BNS events without EM counterparts improves the measurability of $\Xi_0$ from the case with EM counterparts alone by a factor of a few. Notice also that for the combined case, BNSs with EM counterparts have a noticeable contribution when $\alpha \gtrsim 0.1$ for ET alone and $\alpha \gtrsim 0.01$ for multi-band observations (where the red curves drop). Furthermore, when $\alpha \sim 1$ (i.e.~most of BNSs have EM counterparts), multi-band observations significantly improve the bound on $\Xi_0$ from the case with ET alone. This is because when $\alpha \sim 1$, the error budget in the luminosity distance measurement is different between ET and multi-band cases as already explained in Sec.~\ref{sec:redshift_inference} and in Fig.~\ref{fig:ErrorCompare}. 

\begin{figure}[htp]
\includegraphics[width=8.5cm]{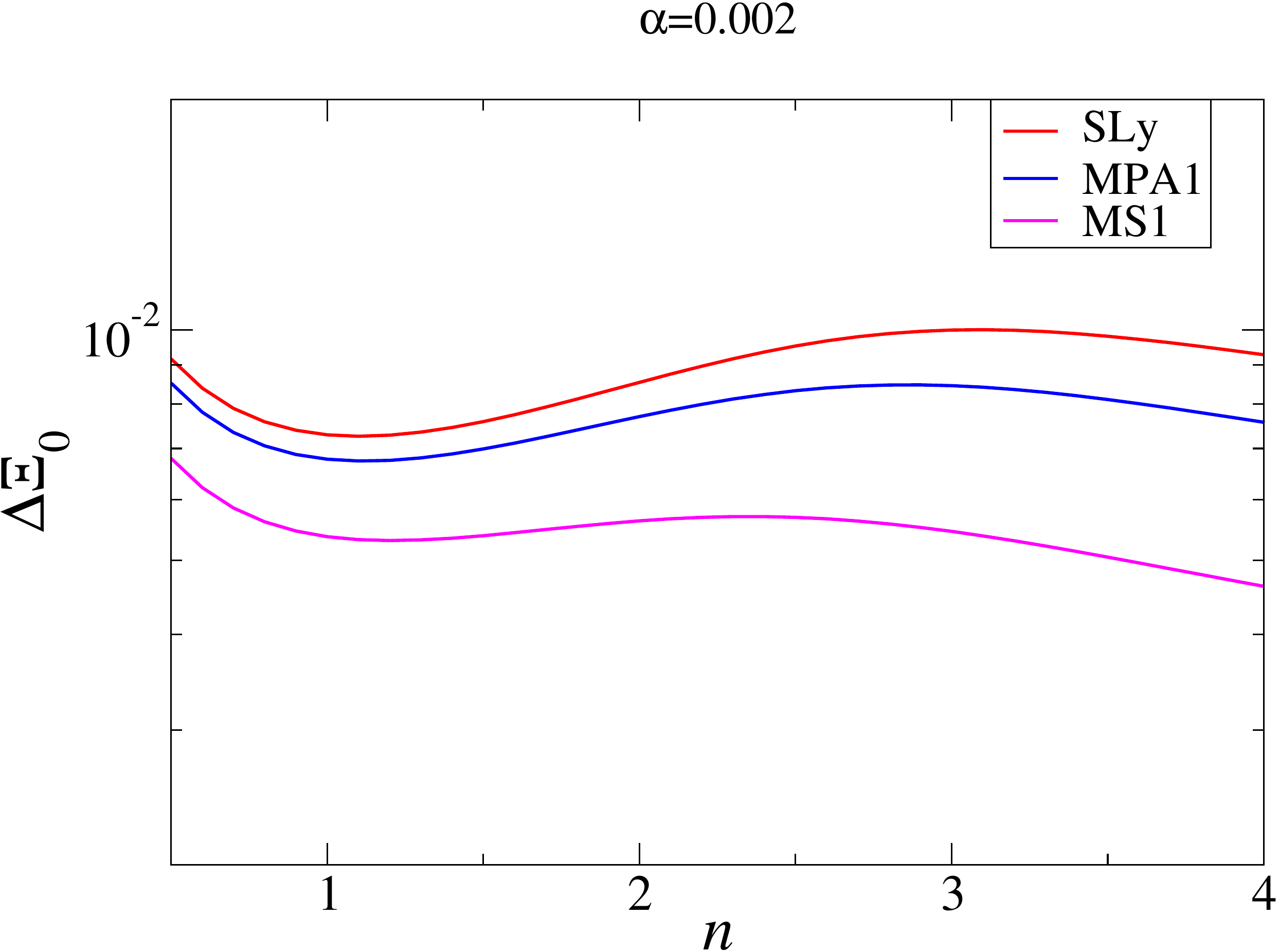}
\caption{
Measurability of $\Xi_0$ against the index $n$ in the modified GW luminosity distance in Eq.~\eqref{eq:mod GR} for three representative EOSs. We consider a multi-band observation with combined BNS events with and without EM counterparts. We fix the fraction of BNSs with EM counterparts as $\alpha = 2\times 10^{-3}$. The observational time is 3 years. Observe that the measurability is not very sensitive to the choice of $n$. 
}
\label{fig:Result 4}
\end{figure}

Next, Fig.~\ref{fig:Result 4} presents the measurability of $\Xi_0$ against the index $n$ in the luminosity distance ratio expression (Eq.~\eqref{eq:mod GR}) for a multi-band observation with combined BNS events (both with and without redshift identification through EM counterparts) for $\alpha=2\times 10^{-3}$ \footnote{The fraction $\alpha = 2 \times 10^{-3}$ is derived for short gamma-ray bursts assuming that 2\% of them points to us and only 10\% of them can have measurable redshift due to noisy spectrum, dimming at high redshift, etc.~\cite{Nishizawa_2012}. This fraction can be larger for other sources, such as kilonova, or if we take into account off-axis emission.}. We show the results for the three representative EOSs. Notice first that the measurement error of $\Xi_0$ is mostly insensitive to $n$ and varies only by $\sim 20\%$. Notice also that the error decreases for stiffer EOSs (MS1), which is consistent with the measurement error of $z$ in Fig.~\ref{fig:Redshift}.

\subsection{Mapping to Horndeski Theories}
\label{sec:mapping_bounds}

\begin{figure}[htp]
\includegraphics[width=8.5cm]{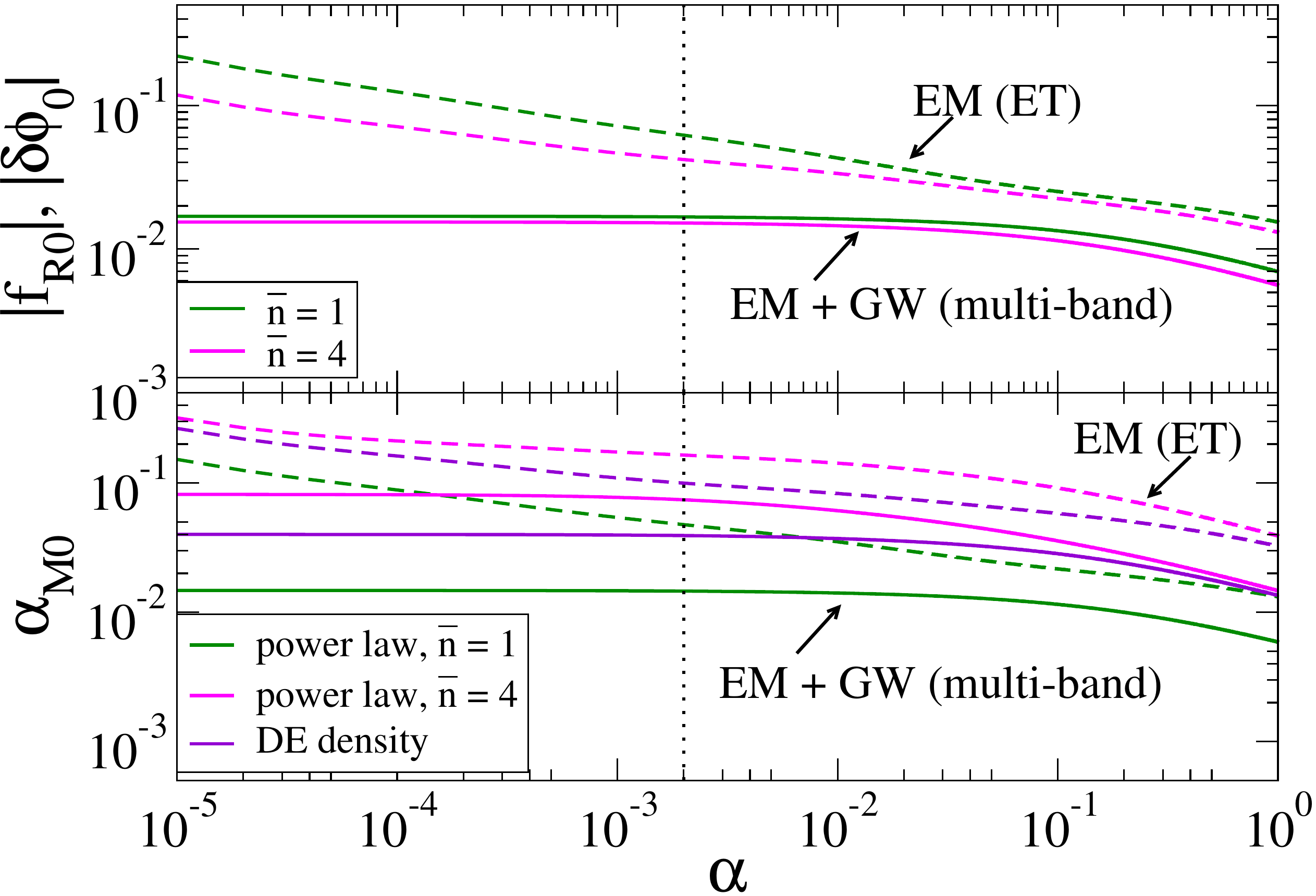}
\caption{
(Top) Projected bounds on parameters in the HS $f(R)$ gravity ($|f_{R0}|$) and Brans-Dicke theory ($|\delta \phi_0|$)  as a function of the fraction $\alpha$ for redshift identification through EM counterparts. We present the bounds for the following two cases: (i) GW observations of BNSs with ET where the sources' redshifts are identified through EM counterparts only (dashed), and (ii) multi-band GW observations of BNSs whose redshifts are identified through either  EM counterparts or GW tidal effects (solid). For each case, we show the bounds for two choices of the positive index $\bar n$. We assume 3-year observations and the SLy EOS.  The dashed vertical line at $\alpha = 2 \times 10^{-3}$~\cite{Nishizawa_2012} corresponds to an example value for the fraction of BNSs with redshift identification.
(Bottom) A similar bound on  the phenomenological $\alpha_\mathrm{M}$ models in Eqs.~\eqref{eq:alphaM_powerlaw} and~\eqref{eq:alphaM_DE}.
}
\label{fig:HS&alpha}
\end{figure}

Finally, we consider mapping the bounds on the modified GW propagation parameter $\Xi_0$ to those on scalar-tensor theories and phenomenological models. The top panel of Fig.~\ref{fig:HS&alpha} shows bounds on the HS $f(R)$ gravity $|f_{R0}|$ and Brans-Dicke theory $|\delta \phi_0|$ as a function of $\alpha$ for various choices of the positive integer $\bar n$. 
Observe that the addition of BNSs with redshift identification through tidal measurements and the use of multi-band observations improve the bounds on these theories from the case with ET observations of BNSs with EM counterparts by a factor of 2--10. Observe also that the bounds are insensitive to a variation in $\bar n$, especially for the multi-band case. 

Similarly, the bottom panel of Fig.~\ref{fig:HS&alpha} presents bounds on $\alpha_{M0}$ in the two phenomenological models mentioned in Sec.~\ref{sec:phenom_alphaM}. Notice that the amount of improvement on the bounds with the addition of BNSs without EM counterparts and multi-band observations is similar to those on scalar-tensor theories in Fig.~\ref{fig:HS&alpha}.
Notice also that the variation in $\bar n$ is larger for this case than that for scalar-tensor theories in the top panel. 

\section{\label{sec:Conclusion}Conclusions and Discussions}

In this paper, we considered using GWs from BNS mergers both with and \emph{without} EM counterparts to probe a modified GW propagation effect in the amplitude due to a modified friction in the tensor perturbation evolution. For the events without EM counterparts, we use the tidal information to break the degeneracy between the redshift and the mass~\cite{Messenger_2012}. We found that by including BNSs without EM counterparts and using multi-band GW observations between ET and DECIGO, one can improve the measurability on the modified GW propagation parameter $\Xi_0$ by a factor of a few compared to the case with ET observations of BNSs with EM counterparts that has been studied previously. We further mapped these projected bounds on $\Xi_0$ to those on specific non-GR theories and phenomenological models. For example, we found that a parameter in an $f(R)$ gravity can be constrained to $|f_{R0}| \lesssim 10^{-2}$. These findings show the impact of using the tidal information and multi-band observations to probe a modified GW propagation (or modified friction) effect entering in the waveform amplitude. 

We end by presenting possible directions for future avenues. One could improve the analysis here by carrying out a Bayesian parameter estimation study (instead of a Fisher analysis) and drawing BNSs from a population model to allow for different parameters (like masses). 
One should also relax the sky-averaged assumption and account for sky location and  orientation of a BNS. This could be important given that there was a strong correlation between the luminosity distance and the inclination angle for GW170817~\cite{Abbott:2018wiz}. However, in Appendix~\ref{sec:deg}, we carried out an additional analysis by relaxing the sky-averaged assumption for DECIGO and showed that in most cases, the measurement error for the luminosity distance is still smaller than that from the redshift measurement. This suggests that the result presented here with the sky-averaged analysis should not change much for multi-band observations even if one accounts for the correlation. It would be also important to  take into account systematic uncertainties due to imperfect knowledge of the EOS and 
certain universal relations may help to break the degeneracy among various tidal parameters~\cite{Yagi:2015pkc,Yagi:2016qmr,Yagi:2016bkt,Chatziioannou:2018vzf,Abbott:2018exr}. Lastly, one could also attempt to combine the tidal method presented here with other approaches that do not require EM counterparts, such as correlating dark sirens with galaxy catalogs~\cite{DelPozzo:2011yh,Abbott:2019yzh,finke2021cosmology} or using the known NS mass distribution~\cite{Taylor:2011fs,Taylor:2012db}. 

\acknowledgments

N.J. and K.Y. acknowledge support from the Owens Family Foundation. K.Y. also 
acknowledges support from
NSF Grant PHY-1806776, NASA Grant 80NSSC20K0523, and a Sloan Foundation Research Fellowship. 
K.Y. would like to also thank the support by the COST Action GWverse CA16104 and JSPS KAKENHI Grants No. JP17H06358.

\appendix

\section{\label{sec:Appendix}Additional Scalar-tensor Theory and Phenomenological Model}

\begin{figure}[t!]
\includegraphics[width=8.5cm]{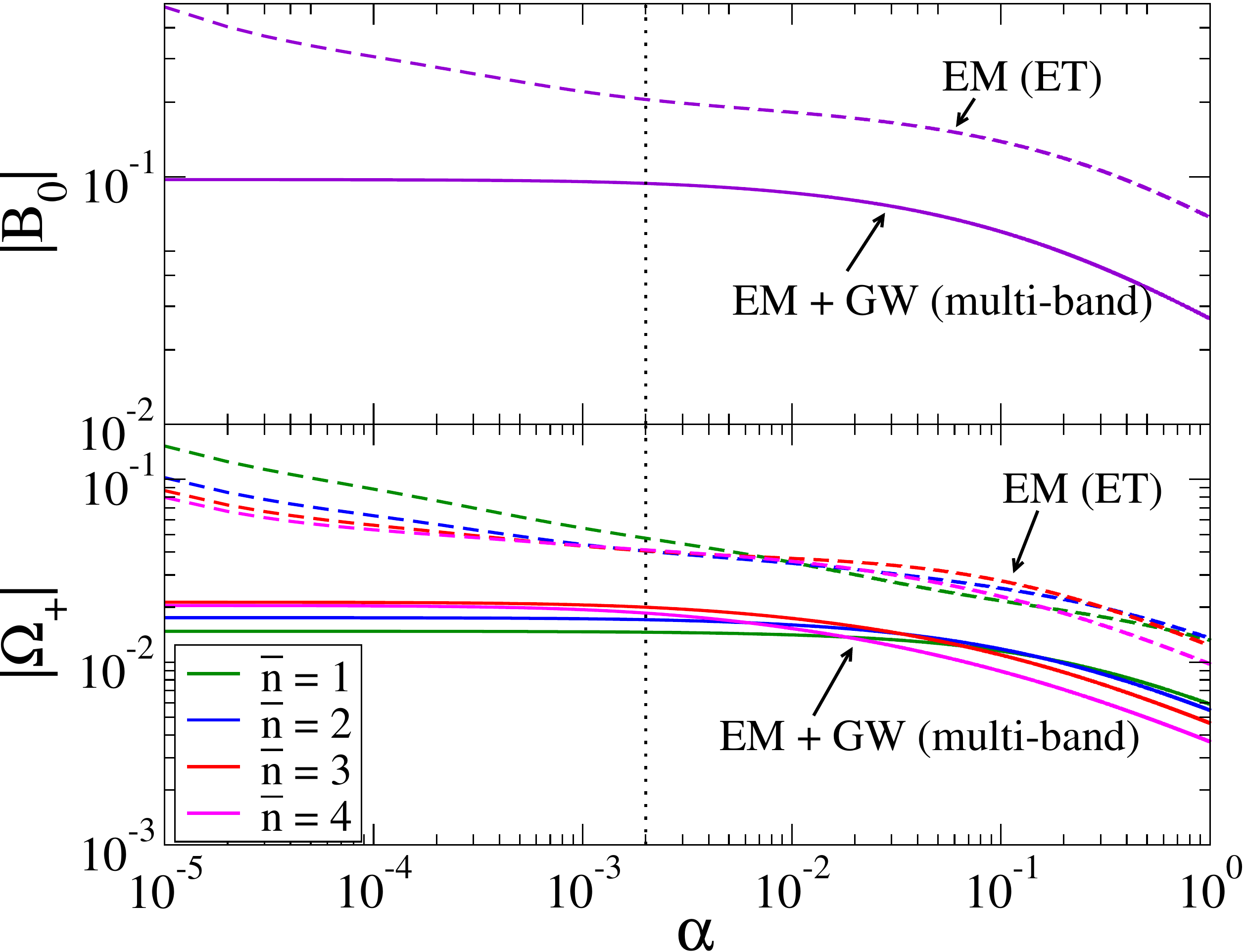}
\caption{
Similar to Fig.~\ref{fig:HS&alpha} but for the bounds on the Compton wavelength parameter $|B_0|$ in the designer $f(R)$ gravity (top) and $|\Omega_+|$ in the power law $M_\mathrm{eff}$ formalism (bottom). 
}
\label{fig:Designer}
\end{figure}

In this appendix, we present the mapping between the modified GW propagation parameters $(\Xi_0,n)$ to additional scalar-tensor theories and phenomenological models, and present future projected bounds on these theories/models through tidal measurement of BNS mergers. 
The mapping is summarized in Table~\ref{tab:mapping}.

\begin{itemize}

\item \emph{Designer $f(R)$ gravity}~\cite{Song_2007}: Other than the HS model, an interesting $f(R)$ gravity model includes the designer model that exactly reproduces the standard cosmological expansion history. The model is characterized by the Compton wavelength parameter
\begin{eqnarray}
B_0 &\equiv&  \frac{H f'_R}{H'(1+f_R)} \bigg|_{0}
 \approx -2.1 \Omega_M^{-0.76} f_{R0}\,.
\end{eqnarray}
The top panel of Fig.~\ref{fig:Designer} presents the bound on $|B_0|$ with GWs from BNSs using a three-year observation of a multi-band network as a function of $\alpha$. We used $n = 2.34$, which is close to $n=2.5$ in Fig.~\ref{fig:Result 1} and thus follows the same trend. Observe that the bounds on $B_0$ increases by a factor of 2 -- 5 if we add BNS events without EM counterparts. 
\item  \emph{power law $M_\mathrm{eff}$}:
On top of the phenomenological models for $\alpha_M$, we consider a phenomenological model on the effective Planck mass $M_\mathrm{eff}$. As an example, we consider a simple power law model for $M_{\mathrm{eff}}^2$ given by~\cite{Lombriser_2016} 
\begin{equation}
\label{eq:M_eff_power_law}
M_\mathrm{eff}^2 = \frac{1}{8\pi} \left( 1 + \Omega_+ a^{\bar n}  \right)\,,
\end{equation}
where $\Omega_+$ and $\bar n$ are constant parameters. $\alpha_M$ in this model is given by
\begin{equation}
\alpha_M = \frac{\bar{n} \Omega_+ a^{\bar{n}-1}}{1 + \Omega_+ a^{\bar n}}\,.
\end{equation} 
Using the mapping in Table~\ref{tab:mapping}, we present in the bottom panel of Fig.~\ref{fig:Designer} the projected bounds on $|\Omega_+|$ for BNSs with and without EM counterparts for various $\bar n$. Observe that the addition of BNSs without EM counterparts improve the bound by an order of magnitude for small $\alpha$ and $\bar n$. On the other hand, the improvement is by a factor of a few irrespective of $\bar n$ when $\alpha \sim 1$.

\end{itemize}

\section{\label{sec:Appendix2}Observation time and EOS dependence on $\Delta \Xi_{0}$}

\begin{figure}[htb!]
\includegraphics[width=8.5cm]{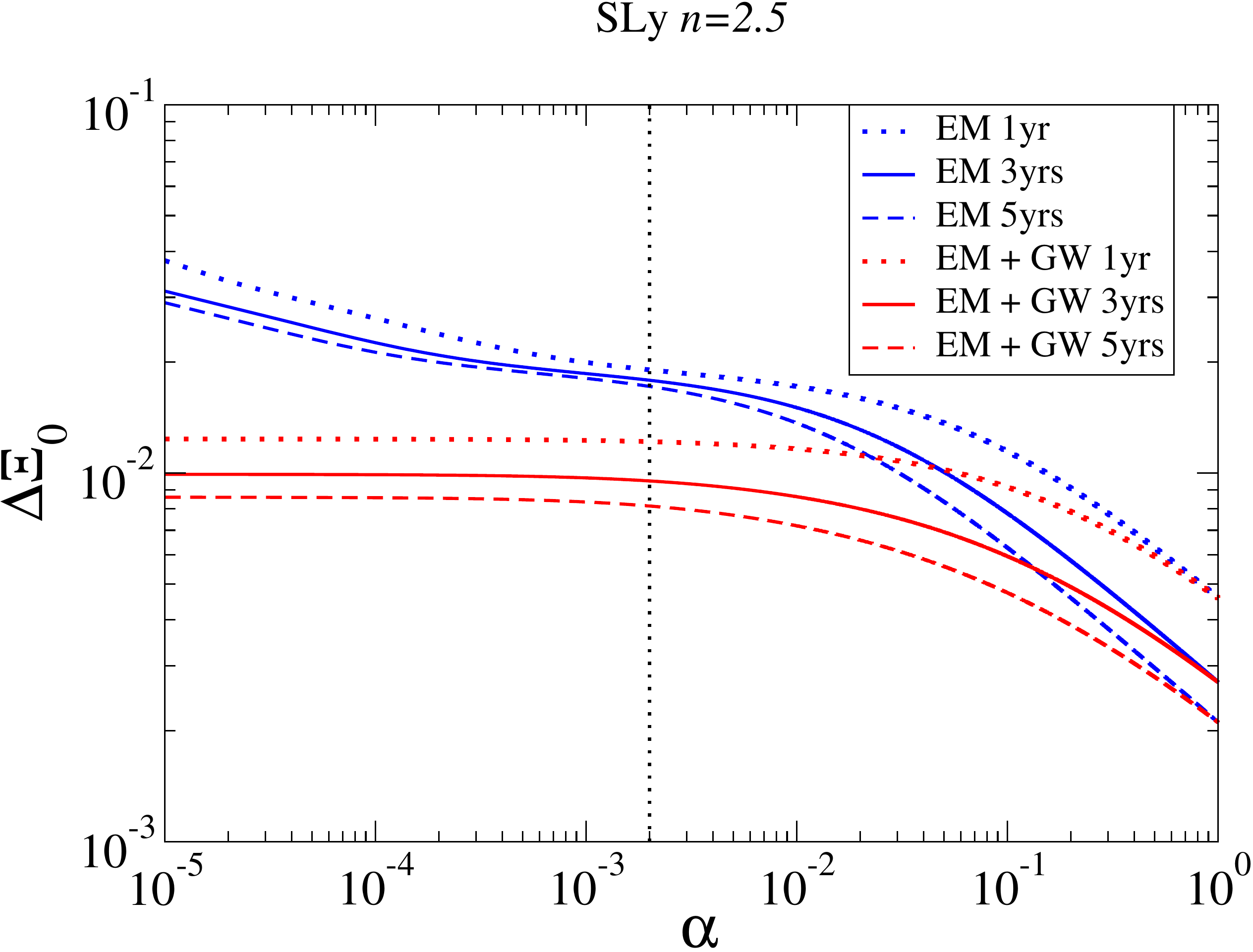}
\caption{
Measurement errors on the modified GW propagation parameter $\Xi_0$ as a function of $\alpha$ for three different observation periods. We consider BNSs with redshift identified from EM counterparts only (blue), as well as those with redshift identification by EM and GW observations (red). We fix $n=2.5$, choose the SLy EOS and consider multi-band observations.
}
\label{fig:Result 2}
\end{figure}

In this appendix, we carry out some additional investigations on the measurability of $\Xi_0$ with multi-band GW observations. Figure~\ref{fig:Result 2} presents how $\Delta \Xi_0$ depends on the observation period. Notice that the observation time has the most significant effect when $\alpha \sim 1$. For this case, the error on the luminosity distance measurement is dominated by the lensing that is independent of the observation time. Moreover, 
the prior on the second Fisher matrix $\tilde F_{ij}$ in Eq.~\eqref{eq:Fisher_gw+em} is less important and the measurability scales with $T_\mathrm{obs}^{-1/2}$ since the number of BNS events increases linearly with $T_\mathrm{obs}$ (see Eq.~\eqref{eq:f(z)}). On the other hand, for smaller $\alpha$, the prior on $\tilde F_{ij}$ becomes more important and the above scaling breaks down. Notice also that the observation time has a larger effect on the case with all BNSs (with and without EM counterparts) than BNSs with EM counterparts only. This is because for the former, the error on the luminosity distance measurement is dominated by the redshift uncertainty, and a longer observation time helps more to break the degeneracy between the redshift and other parameters.

\begin{figure}[htb!]
\includegraphics[width=8.5cm]{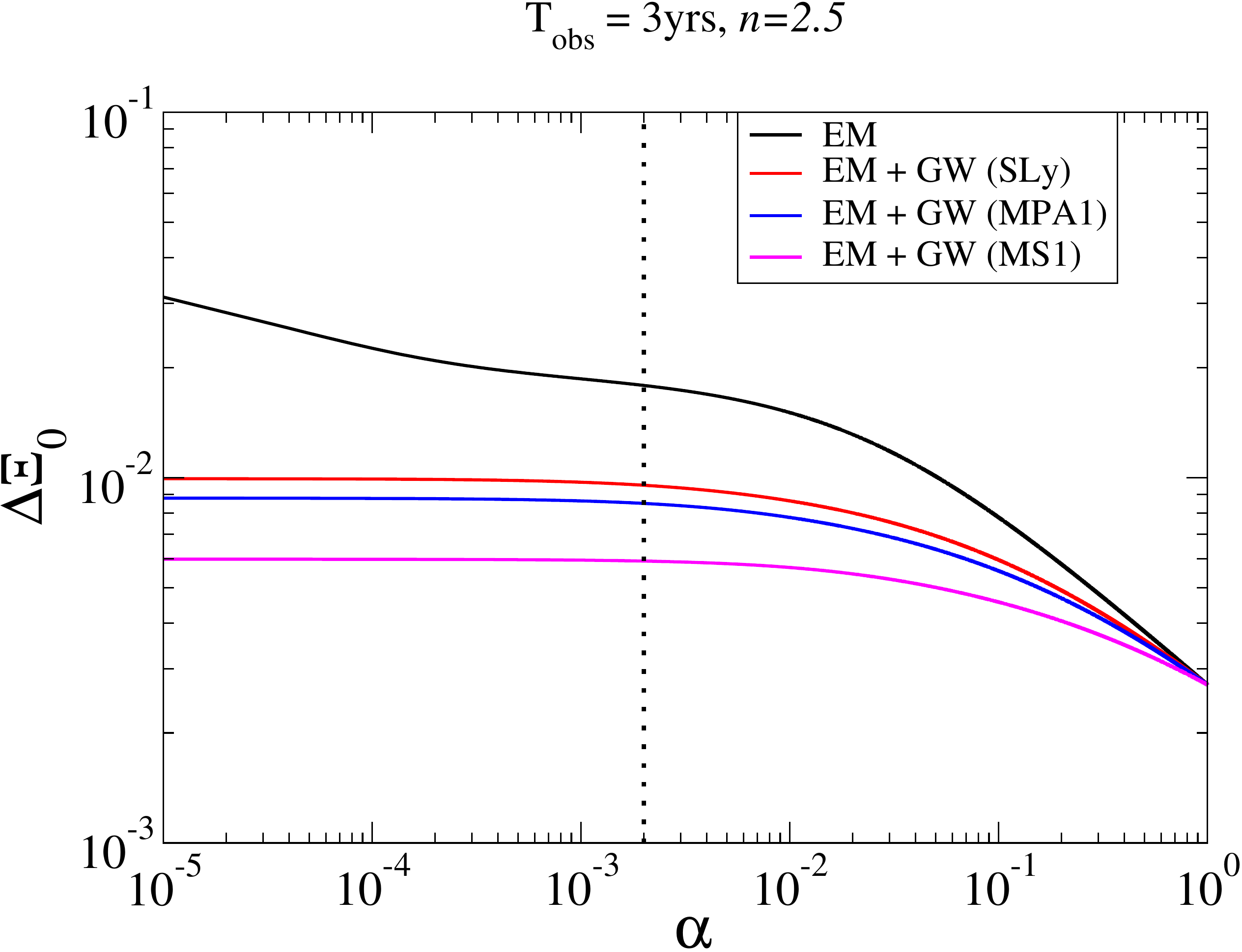}
\caption{
Similar to Fig.~\ref{fig:Result 2} but showing how $\Delta \Xi_0$ varies with EOSs. We fix $T_\mathrm{obs}=3$yrs and $n=2.5$.
}
\label{fig:Result 3}
\end{figure}

Figure~\ref{fig:Result 3} presents $\Delta \Xi_0$ with multi-band observations for the three representative EOSs. For the case with BNSs with EM counterparts alone, EOS only affects the first Fisher matrix $\tilde \Gamma_{ij}$ through the maximum frequency cutoff. Since the effect is small, we only consider the SLy EOS for this case. Notice that the measurability of $\Xi_0$ improves as we make the EOS stiffer. This is as expected from the measurability of the redshift from Fig.~\ref{fig:Redshift}.

\section{\label{sec:Appendix1}Inclusion of $\tilde{\lambda}_0$ and $\tilde{\lambda}_1$ }

\renewcommand{\arraystretch}{1.5}
\begin{table}[]
\begin{adjustbox}{max width=\textwidth}
\begin{tabular}{c||c c|c c}
\multirow{2}{*}{} & \multicolumn{2}{c|}{SLy} & \multicolumn{2}{c}{MS1} \\ 
 & \multicolumn{1}{l}{\begin{tabular}[c]{@{}l@{}}pessimistic \\ (30 BNSs)\end{tabular}} & \multicolumn{1}{l|}{\begin{tabular}[c]{@{}l@{}}optimistic\\ (384 BNSs)\end{tabular}} & \multicolumn{1}{l}{\begin{tabular}[c]{@{}l@{}}pessimistic \\ (30 BNSs)\end{tabular}} & \multicolumn{1}{l}{\begin{tabular}[c]{@{}l@{}}optimistic\\ (384 BNSs)\end{tabular}} \\ \hline \hline
$\tilde{\lambda}_0$ & \multicolumn{2}{c|}{4.46} & \multicolumn{2}{c}{12.41} \\ 
$\sigma_{\tilde{\lambda}_0}$ & 0.039 & 0.014 & 0.029 & 0.009 \\ \hline
$\tilde{\lambda}_1$ & \multicolumn{2}{c|}{-1.99} & \multicolumn{2}{c}{-3.35} \\ 
$\sigma_{\tilde{\lambda}_1}$ & 0.025 & 0.0125 & 0.019 & 0.009 \\ 
\end{tabular}
\end{adjustbox}
\caption{Values of $\tilde{\lambda}_0~[10^{36} \mathrm{g} \mathrm{cm}^2\mathrm{s}^2]$, $\tilde{\lambda}_1~[(10^{36} \mathrm{g} \mathrm{cm}^2\mathrm{s}^2/M_\odot)]$ and their standard deviations for Gaussian priors for SLy and MS1. The priors are taken from the results of two cases under the detection of HLV in Section 5.4 in \cite{Wang_2020}, the pessimistic case with 30 BNSs and the optimistic case with 384 BNSs.}
\label{tab:lambdapriors}
\end{table}

In this appendix, we study how the imperfect knowledge of the EOS may affect the measurability of the redshift. For this, we include $\tilde \lambda_0$ and $\tilde \lambda_1$
 into a search parameter set $\theta^i$ in Eq.~\eqref{eq:first Fisher} for the first Fisher analysis: 
 \begin{eqnarray}
\label{eq:first Fisher prior}
	\theta^i = \left(\mathrm{ln}\mathcal{M}_z, \eta,t_c,\phi_c,\mathrm{ln}A,\mathrm{ln}\tilde{\lambda}_0,\mathrm{ln} \tilde{\lambda}_1,\mathrm{ln}z \right).
\end{eqnarray}
 For simplicity, we follow~\cite{Cutler:1994ys,PhysRevD.52.848} and assume a Gaussian prior with standard deviations $\sigma_{\tilde \lambda_0}$ and $\sigma_{\tilde \lambda_1}$. The effective Fisher matrix now becomes
 \begin{equation}
\tilde{\Gamma}_{ij} = \sum_{A} \Gamma_{ij}^{(A)} 
+ \frac{\delta_{ij}}{\left(\sigma_{\theta^i} \right)^2}.
\end{equation}

To give an example, we consider a prior for $\tilde \lambda_0$ and $\tilde \lambda_1$ that corresponds to measuring them through a network of LIGO Hanford/Livingston and Virgo (HLV) shown in Table~\ref{tab:lambdapriors} that is taken from~\cite{Wang_2020}. Following this reference, we assume that all BNSs with $z < 0.1$ detected through such a network has EM counterparts and can be used to measure $\tilde \lambda_0$ and $\tilde \lambda_1$. This is somewhat optimistic, though the authors in~\cite{Wang_2020} found that the measurability of these tidal parameters do not change much even if one only uses BNSs with $z < 0.05$. 
 
 \begin{figure}[thb]
\includegraphics[width=8.5cm]{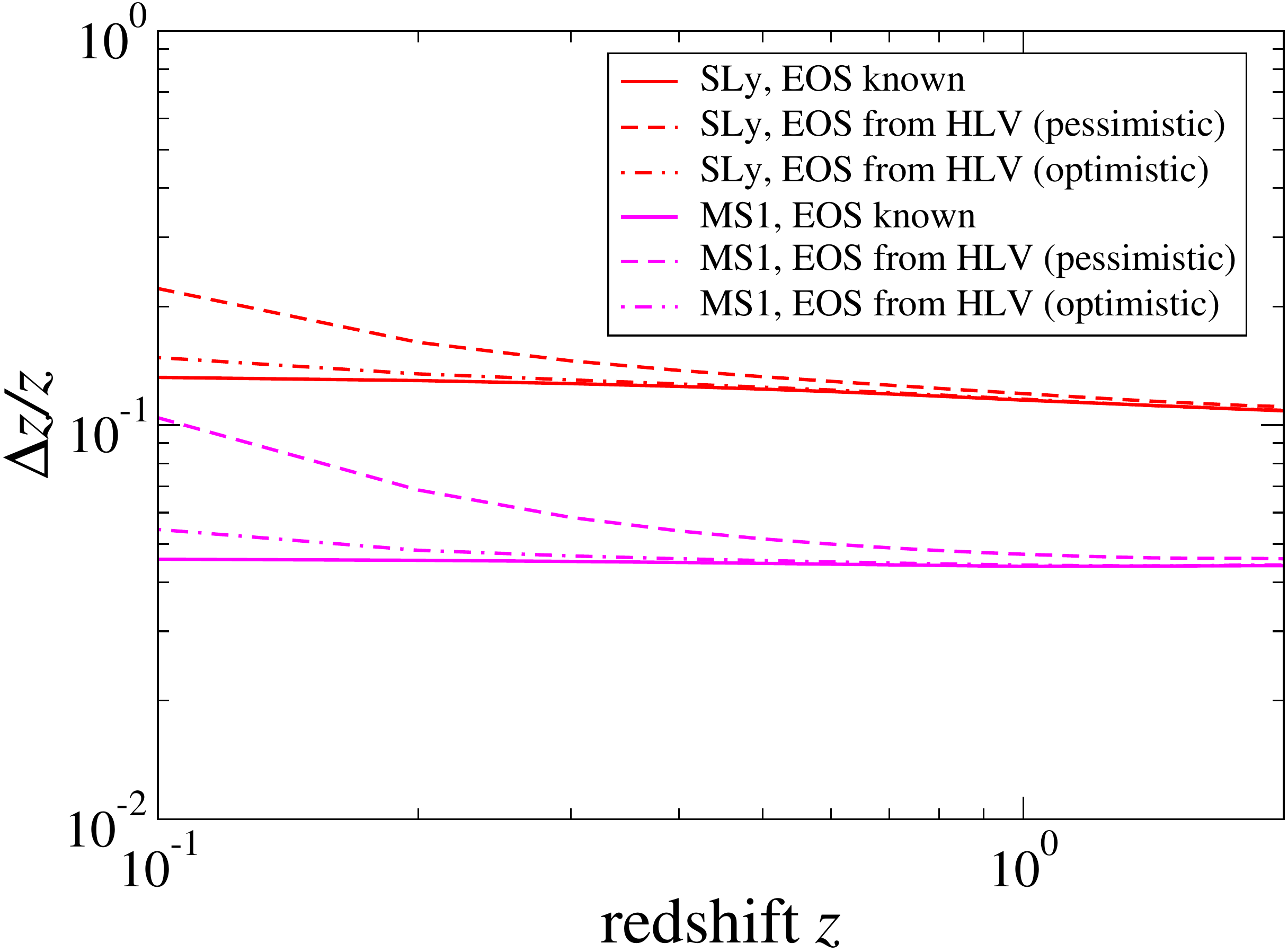}
\caption{
The measurability of the redshift with multi-band GW observations for the case where (i)  the EOS is known \emph{a priori} and (ii) $\tilde \lambda_0$ and $\tilde \lambda_1$ are obtained from a network of HLV observations~\cite{Wang_2020}. For the latter, we consider both pessimistic (30 BNSs) and optimistic cases (384 BNSs) for 3-yr observations. We show the results for SLy (soft) and MS1 (stiff) EOS.
}
\label{fig: have prior}
\end{figure}
 
 Figure~\ref{fig: have prior} presents the measurability of the redshift for multi-band GW observations where $\tilde \lambda_0$ and $\tilde \lambda_1$ are included in the search parameter for Fisher analyses for the SLy and MS1 EOSs. We consider a pessimistic (optimistic) case with 30 (384) detected BNSs with $z < 0.1$ for a 3-yr observation. For reference, we show the result without $\tilde \lambda_0$ and $\tilde \lambda_1$ in the search parameter set from Fig.~\ref{fig:Redshift}. Notice that the uncertainty in the EOS affects the measurability of the redshift only for BNSs with  low $z$. Moreover, such an uncertainty on the EOS will be reduced  if one uses ET instead of LHV. We thus expect the effect of imperfect knowledge of the EOS to be small and neglect them in the main text.

\section{Degeneracy between luminosity distance and binary orientation}
 \label{sec:deg}

 In this appendix, we estimate the amount of degeneracy between the luminosity distance and binary orientation for multi-band observations. Since the measurability of the luminosity distance for the multi-band observation is mostly determined by observations with DECIGO (due to its high SNR and a large effective baseline of 1AU), we focus on the latter for simplicity. The binary inclination varies over time due to the motion of DECIGO, and thus it is useful to work in a barycentric frame (centered at the Sun)~\cite{Cutler:1997ta,Berti_2005,Yagi:2009zm,Yagi:2009zz}. In such a frame, we can describe the sky location of a BNS by $(\theta_s, \phi_s)$ and the direction of its orbital angular momentum as $(\theta_L, \phi_L)$. Following~\cite{Yagi:2009zz}, we perform a new Fisher analysis with search parameters given by\footnote{In this appendix, we do not include $z$ since we focus on DECIGO which is insensitive to the effect close to merger. This does not affect the luminosity distance measurement since the amplitude parameters are mostly uncorrelated with the phase parameters.}
 \begin{equation}
 \theta^i = \left(\mathrm{ln}\mathcal{M}_z, \eta,t_c,\phi_c,\mathrm{ln}A,\theta_s,\phi_s,\theta_L,\phi_L \right),
 \end{equation}
 and we take into account the motion of the detectors.
 We use a restricted post-Newtonian waveform where we only consider the leading Newtonian contribution for the amplitude while we include up to 2PN order in the phase. We carry out a Monte Carlo simulation in which we consider $10^3$ BNSs at $z=1$ with the angle parameters randomly drawn from a uniformly distribution in $\cos \theta_s$, $\phi_s$, $\cos \theta_L$ and $\phi_L$~\cite{Berti_2005,Yagi:2009zm,Yagi:2009zz}.

\begin{figure}[htb]
\includegraphics[width=8.5cm]{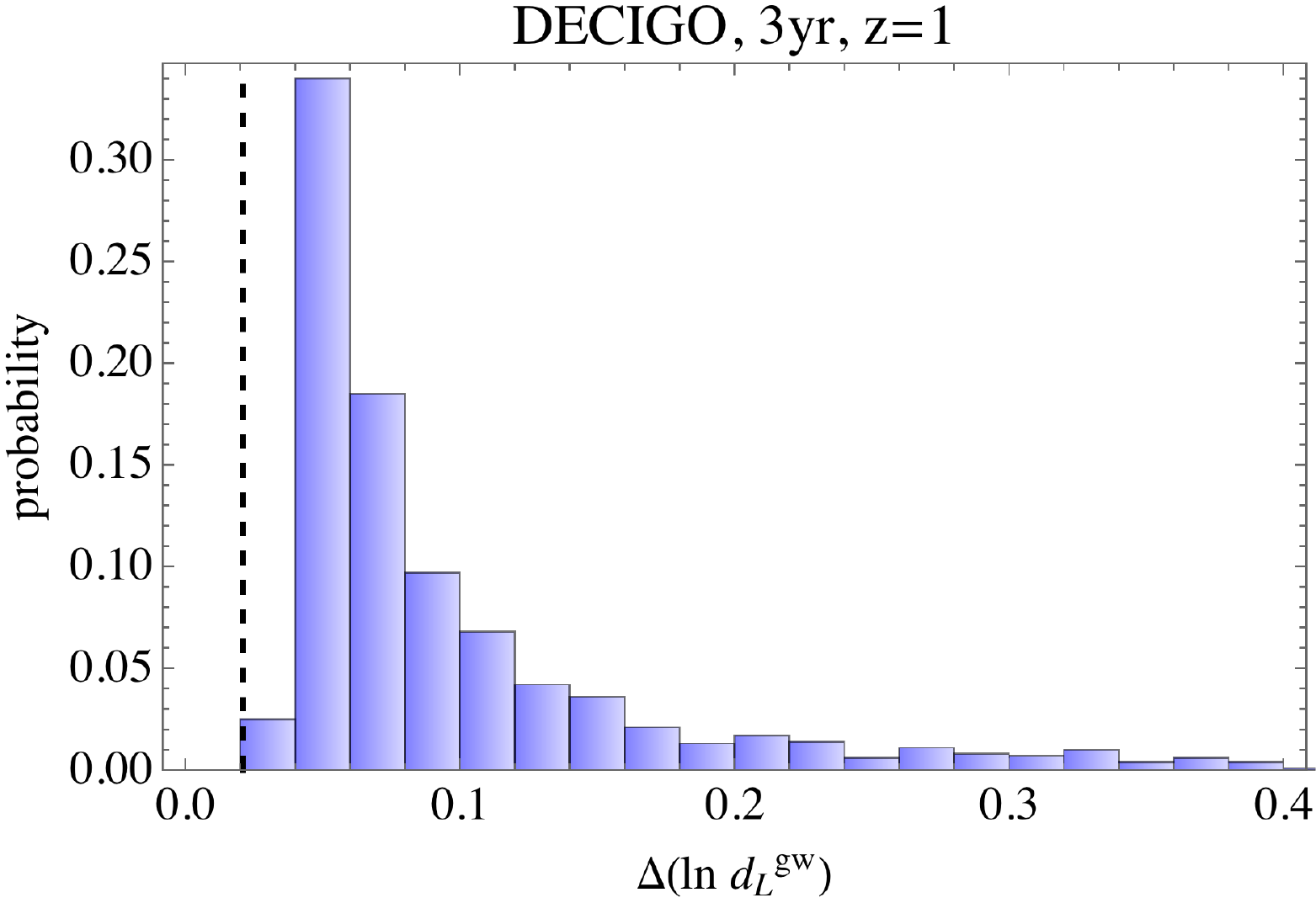}
\caption{Histogram for the probability of the luminosity distance measurability with DECIGO for at $z=1$ whose sky location and orientation are randomly distributed. The black dashed vertical line shows the measurability with the sky-averaged case. Notice that most of binaries have the fractional error of less than 10\% even if we account for the degeneracy between $d_L^\mathrm{gw}$ and binary orientations. 
}
\label{fig:degeneracy}
\end{figure}

 Figure~\ref{fig:degeneracy} presents the distribution of the luminosity distance measurability for a 3-yr observation with DECIGO for BNSs at $z=1$. For comparison, we also show the measurability when we use a sky-averaged waveform as done in the main part of this paper, which roughly agrees with the blue solid  curve in Fig.~\ref{fig:ErrorCompare} at $z=1$ (suggesting that the error is indeed determined from the DECIGO measurement for multi-band observations). Notice that although the sky-averaged analysis underestimates the error, the measurement error is below 10\% for most of BNSs and thus does not exceed the error on the luminosity distance from the redshift measurement. This shows that the bound on $\Xi_0$ for multi-band observations found in this paper through the sky-averaged analysis will not be affected much even if we include the effect of binary sky location and orientation.
 
\bibliography{ref}
%
\end{document}